\documentclass[conference]{IEEEtran}
\usepackage{cryptocode}
\let\labelindent\relax

\usepackage{algorithm,algpseudocode,enumitem}
\usepackage[normalem]{ulem}
\usepackage{booktabs}
\usepackage{multirow}
\usepackage{siunitx}
\usepackage{makecell}
\usepackage{xcolor}
\usepackage{todonotes}
\usepackage{subcaption}

\newcommand{\pprotocol}[5]{{\begin{figure}[#3]
\begin{center}
\fbox{
\hbox{\quad
\begin{minipage}{#4\textwidth}
\small
#5
\end{minipage}
\quad} }
\caption{\label{#2} #1}
\end{center}
\end{figure} } }

\newcommand{\myprotocol}[4]{\pprotocol{#1}{#2}{htp!}{#3}{#4}}

\newcommand{\keygen}{{\sf G}}

\newcommand{\id}{{\sf id}}

\newcommand{\T}{\mathbb{T}}

\newcommand{\ep}{\varepsilon}

\newcommand{\prob}[1]{{\rm Pr}_{#1}}

\newtheorem{defn}{Definition}

\newcolumntype{L}[1]{>{
    \raggedright\let\newline\\\arraybackslash\hspace{0pt}}m{#1}}
\newcolumntype{C}[1]{>{
    \centering\let\newline\\\arraybackslash\hspace{0pt}}m{#1}}
\newcolumntype{R}[1]{>{
    \raggedleft\let\newline\\\arraybackslash\hspace{0pt}}m{#1}}

\newcommand{\IFbox}[2]{{%
    \fbox{\vbox{%
        \center{\textbf{#1}}

        #2%
    }}%
}}

\newcommand{\threenmc}{

\begin{enumerate}[align=left]
    \item[{\bf 1. Initialization:}] The challenger $\mathcal{C}$ initializes several data structures, specified in the next section, to keep track of what transpires in the system during the course of the game. \emph{For example, $\mathcal{C}$ creates sets $V_{\sf corrupt}\subset V$, both  initialized to empty, which keep track of the corrupt vehicles in the system. Or $\mathcal{C}$ can keep track of the corrupt tokens of a vehicle $v$ in the set $T^v_{\sf corrupt}\subset T^v$}.

    \item[{\bf 2. Online Phase:}] The adversary $\mathcal{A}$ interacts with $\mathcal{C}$ using the specified queries, defined later. \emph{For example, if $\mathcal{A}$ wants to add a new vehicle to the system, $\mathcal{A}$ would use the $\text{CreateVehicle}(\cdot)$ query, at which point $\mathcal{C}$ would run Alg 1 and add $(v,EC_v)$ to $V$}. \emph{Or, if $\mathcal{A}$ decides it wishes to attack the anonymity of our scheme, it would issue the $\text{Anonymity}$ query}.

    \item[{\bf 3. The Final Message:}] $\mathcal{A}$ sends its final message and the game ends.  $\mathcal{A}$ wins if it meets one of the winning conditions specified in the next section.
\end{enumerate}
}

\newcommand{\initial}
{
\flushleft
\textbf{Initialization Data:}     
The Long-term keys to the vehicle, pseudonym certificates, and tokens are the critical components of the security model. For these data to be used and presented, we will need algorithms ~\ref{alg:setup}, ~\ref{alg:tokenGen}, and ~\ref{alg:pseudogen}.
\begin{itemize}
        \item[$(1)$] A set $V$ which will contain public/private key pairs $(v,{\sf EC}_{v})$; $V$ represents the set of all cars which ever exist in the system; $V$ is initialized to the emptyset;
        
         \item[$(2)$] A set $V_{\sf corrupt}\subset V$ of corrupt cars; $V_{\sf corrupt}$ is initialized to the emptyset;

         \item[$(3)$] For every $v\in V$, $\mathcal{C}$ keeps track of a set $T^v$ which represents the set of tokens for the vehicle $v$; $\mathcal{C}$ also maintains $T^v_{\sf corrupt}\subset T^v$, the set of corrupted tokens for $v$; every time a new $v$ is added to $V$, $T^v$ and $T^v_{\sf corrupt}$ are initialized to empty sets;
         
         \item[$(4)$] For each $v\in V$, $\mathcal{C}$ keeps track of a set of valid pseudonym certificate $P^v$ at  a given time interval; If the vehicle's PC is exposed, then the same is added to the set $P^v_{\sf corrupt}$;
\end{itemize}

\textbf{Interaction Queries:}
These queries emulate how the adversary might interact with the system in the real world.
\begin{enumerate}
    \item $\text{CreateVehicle}():$ $\mathcal{C} $  runs ~\ref{alg:setup} to generate the pair $(v, \sf EC_v)$ adding it to $V$, $C$; it also creates sets $T^v$ and $T^v_{\sf corrupt}$ which are initialized to empty sets; finally $\mathcal{C}$ sends $v$ to $\mathcal{A}.$ 

    \item $\text{CorruptVehicle}(v):$ $\mathcal{C}$ finds $(v,\sf EC_v)$ in $V$ (if there is no pair in $V$ with first coordinate $v$ then $\mathcal{C}$ does nothing) and returns $\sf EC_v$ to $\mathcal{A}$ and adds the pair to $V_{\sf corrupt}$;

    \item $\text{GetToken}(v):$ $\mathcal{C}$ runs Algorithm ~\ref{alg:tokenGen} with input $v$ and receives a token $\tau$, and adds $\tau$ to the set $T^v$.
    
    \item $\text{CorruptToken(v)}:$ $\mathcal{C}$ finds the token $\tau \in T^v$, adds it to the set $T^v_{\sf corrupt}$ and sends it to the adversary.

    \item $\text{GetPC(v)}:$
    $\mathcal{C}$ searches a valid token $\tau \in T^v$ and runs the Algorithm ~\ref{alg:pseudogen} with input $(v,\tau)$. When it receives a PC, it adds it to the set $P^v$.

    \item $\text{ExposePC(v)}$ The challenger selects a valid pseudonym certificate from the set $P^v$, adds it to $P^v_{\sf corrupt}$, and sends it to $\mathcal{A}$.

\end{enumerate}

\textbf{Challenge Queries:}
These queries indicate that the adversary is ready to break some facet of the scheme's security.

\begin{enumerate}
    \item $\text{Anonymity}(v_0,v_1)$: $\mathcal{C}$ selects valid pseudonym certificates $P^{v_0}$, $P^{v_1}$ for the vehicles $v_0$, and $v_1$ respectively. It then chooses a bit $b$ at random from the set $\{0,1\}$ and sends $P^{v_b}$ to $\mathcal{A}$.
 
    \item $\text{Unlinkability}(v_0,v_1)$: $\mathcal{C}$ selects valid pseudonym certificates from $P^{v_0}$, $P^{v_1}$ for the vehicles $v_0$, and $v_1$ respectively. $\mathcal{C}$ sends them to $\mathcal{A}$. It then chooses a bit $b$ at random from the set $\{0,1\}$ and sends a different PC from the set $P^{v_b}$ to $\mathcal{A}$.

 \item $\text{Forgery}(v)$: $\mathcal{C}$ selects pseudonym certificates from $P^v$ and sends them to the adversary $\mathcal{A}$. These PCs are added to the set $P^v_{\sf corrupt}$.

\end{enumerate}
}

\newcommand{\win}{
\flushleft
 In our security formalization, we give the adversary $\mathcal{A}$ the power to query before and after the challenge message is received from $\mathcal{C}$. To win, the adversary must satisfy one of the conditions below.
\begin{itemize}
    \item  When the adversary sends its final message(the guess bit $b'$) in case of an Anonymity challenge, the probability of the correct message must be $1/2+ non-negl$.

    \item When the adversary sends its final message(the guess bit $b'$) in case of an Unlinkability challenge, the probability of the correct message must be $1/2+ non-negl$.

    \item When the adversary sends its final message, $m$ signed by a PC of the vehicle $v$, in case of a Forgery challenge, the probability of the correct message must be $non-negl$.
\end{itemize}

}

\newcommand{\threenmcbox}{
\begin{figure}[t!]
 \begin{center}\IFbox{}{  \threenmc }\caption{\protect\label{fig:securitydef} VPKI Security Game}
 \end{center}
\end{figure}
}

\newcommand{\initialization}{
\begin{figure}[t!]
 \begin{center}\IFbox{}{  \initial }\caption{\protect\label{fig:initialization} 
 Initialization and Queries.}
 \end{center}
\end{figure}
}

\newcommand{\winning}{
\begin{figure}[t!]
 \begin{center}\IFbox{}{  \win }\caption{\protect\label{fig:winning} Winning Conditions.}
 \end{center}
\end{figure}
}

\graphicspath{ {Images/} }

\newcommand{\ie}{\emph{i.e., }\xspace}
\newcommand{\eg}{\emph{e.g., }\xspace}

\newcommand{\cut}[1]{}

\newcommand{\name}{TVSS}
\newcommand{\namespace}{TVSS }

\newcommand{\nael}[1]{{\color{green} Nael: #1}}

\newcommand{\setup}{{\sf Setup}}
\newcommand{\tokengen}{{\sf TokenGen}}
\newcommand{\pseudogen}{{\sf PseudoGen}}

\newcommand{\revoke}{{\sf Revoke}}

\newcommand{\tw}{{\sf tw}}

\renewcommand{\id}{{\sf id}}

\newcommand{\st}{{\sf st}}
\newcommand{\et}{{\sf et}}
\newcommand{\ltc}{{\text{EC}}}

\renewcommand{\T}{{\sf T}}

\newcommand{\pc}{{PC}}

\newcommand{\TRL}{{\sf TBL}}
\newcommand{\rsu}{\text{RSU}}

\newcommand{\V}{{\sf V}}

\newcommand{\CA}{{\sf CA}}

\newcommand{\BL}{{\sf BL}}
\newcommand{\sk}{{\sf sk}}
\newcommand{\vk}{{\sf vk}}

\newcommand{\msg}{{\sf msg}}

\renewcommand{\keygen}{{\sf KeyGen}}
\newcommand{\sign}{{\sf Sign}}
\newcommand{\encrypt}{{\sf Enc}}
\newcommand{\decrypt}{{\sf Dec}}
\newcommand{\verify}{{\sf Verify}}

\newcommand{\ct}{{\sf ct}}
\newcommand{\calC}{\mathcal{C}}
\newcommand{\enckey}{{\sf ek}}
\newcommand{\deckey}{{\sf dk}}

\newcommand{\revoked}{{\sf revoked}}
\newcommand{\hastokens}{{\sf hasTokens}}
\newcommand{\timevalid}{{\sf timeValid}}

\newcommand{\deciderevoke}{{\sf decideRevoke}}

\newcommand{\gettoken}{{\sf getToken}}
\newcommand{\update}{{\sf Update}}
\newcommand{\findltc}{{\sf findLTC}}

\newcommand{\calA}{\mathcal{A}}
\renewcommand{\calC}{\mathcal{C}}

\ifCLASSINFOpdf
\else
\fi
\usepackage{url}

\begin{document}
%
\title{Token-based Vehicular Security System (\name): Scalable, Secure, Low-latency Public Key Infrastructure for Connected Vehicles}



%

\author{\IEEEauthorblockN{Abdulrahman Bin Rabiah\IEEEauthorrefmark{1}\IEEEauthorrefmark{3},
Anas Alsoliman\IEEEauthorrefmark{2},
Yugarshi Shashwat\IEEEauthorrefmark{1},
Silas Richelson\IEEEauthorrefmark{1} and Nael Abu-Ghazaleh\IEEEauthorrefmark{1}}
\IEEEauthorblockA{
\IEEEauthorrefmark{1}Department of Computer Science and Engineering, University of California, Riverside\\
\IEEEauthorrefmark{2}Donald Bren School of Information \& Computer Sciences, University of California, Irvine\\
\IEEEauthorrefmark{3}College of Computer and Information Sciences, King Saud University\\
Email: abinr001@ucr.edu, aalsolim@uci.edu, yshas001@ucr.edu,  \{silas, nael\}@cs.ucr.edu}
}



\maketitle

\begin{abstract}
Connected and Autonomous vehicles stand to drastically improve the safety and efficiency of the transportation system in the near future, while also reducing pollution.   These systems leverage communication to coordinate among vehicles and infrastructure in service of a number of safety and efficiency driver assist, and even fully autonomous applications.  Attackers can compromise these systems in a number of ways including by falsifying communication messages, making it critical to support security mechanisms that can operate and scale in dynamic scenarios.  Towards this end, we present \name, a new VPKI system which improves drastically over prior work in the area (including over SCMS; the US department of transportation standard for VPKI).  \name\ leverages the idea of unforgeable tokens to enable rapid verification at the road side units (RSUs), which are part of the road infrastructure at the edge of the network.  This edge based solution enables agile authentication by avoiding the need for back-end servers during the potentially short contact time between a moving vehicle and the infrastructure.  It also results in several security advantages: (1) {\em Scalable Revocation:} it greatly simplifies the revocation problem, a difficult problem in large scale certificate systems; and (2) {\em Faster Refresh:} Vehicles interact more frequently with the system to refresh their credentials, improving the privacy of the system.  
We provide a construction of the system and formally prove its security.  Field experiments on a test-bed we develop consisting of on-board units (OBUs) and RSUs shows substantial reduction in the latency of refreshing credentials compared to SCMS, allowing the system to work even with smaller window of connectivity when vehicles are moving at higher speeds.  Notably, we are able to execute the bottleneck operation of our scheme with a stationary RSU while traveling at highway speeds (more than three times faster than the maximum speeds supported by SCMS).  
\end{abstract}



%

\section{Introduction}
\label{sec:intro}


Autonomous and Connected vehicles act as a coordinated ensemble rather than individual vehicles, coordinating through wireless communication with each other and the infrastructure.
The US Department of Transportation (USDOT)\cite{ITSJointProgram} has identified different CV applications, including improved safety, traffic flow and efficiency, and reduced environmental footprints such as fuel consumption and emissions. Safety applications such as \emph{Basic Safety Messages} are projected to reduce road fatalities by 80\%, a decrease of roughly 30,000 per year~\cite{najm2010frequency,facts2018motor}.  More than 50 intersections in Pittsburgh, PA were fitted with intelligent traffic signal control systems between 2012 to 2016, which reduced travel times through these intersections by 26\%~\cite{promises2022pittsburgh}.  Coordinated driving applications such as \emph{Cooperative Adaptive Cruise Control} allow connected vehicles to travel closely in a convoy/platoon, reducing aerodynamic drag, improving fuel efficiency, and lowering the carbon footprint of the entire transportation system by a projected $15\%$~\cite{barth2015intelligent}.
 Commercially, a number of companies such as Waymo and Cruise have started offering self-driving rides in different cities in the US and across the world.  It is estimated that by 2035, autonomous driving will generate between 300 to 400 billion USD of revenue~\cite{Deichmann2023AD}.

As Connected and Autonomous vehicular systems move from small scale experiments, to wide deployment, it is critical to anticipate potential security and privacy challenges~\cite{thing2016autonomous,abdo2019application,sun2021survey,pham2020survey}.  
Security in these types of transportation applications is extremely important as a breach could cause accidents or otherwise disrupt the flow of traffic~\cite{abdo2019application}.   There are a number of threats that can arise in such large scale and dynamic system open to participation of potentially malicious actors.   It is important to provide facilities within the system to enable building secure applications that resist attacks.  At the core of building a secure system is to make sure that the communication among the vehicles and the infrastructure is protected using a Vehicular Public Key Infrastructure (VPKI) empowering participants with cryptographic tools to support authentication, integrity and confidentiality.   A VPKI differs from conventional PKIs in the scale of the system, the presence of mobility and the need to protect the anonymity of the cars by using pseudo-credentials. 

 Designing such a system requires addressing some non-standard security issues having to do with the fact that a vehicle's identity and position over time are sensitive information.  
Indeed, significant prior work on this topic has culminated in the Security Certificate Management System (SCMS)~\cite{USDOTSCMS} which has been adopted by the US Department of Transportation (a similar European standard is also proposed by ETSI~\cite{etsi2009intelligent}).
   These standards provide a public key infrastructure (PKI) for vehicles to use to authenticate themselves which promises a standard unforgeability security guarantee as well as two additional security features called \emph{anonymity} and \emph{unlinkability}.  Roughly speaking, anonymity says that a vehicle should be able to authenticate itself without revealing its long-term vehicle identity; unlinkability demands that it should not be possible to identify the same vehicle authenticating itself in two different time periods.  So if Alice's car authenticates itself today on highway 1, anonymity guarantees that no one can deduce ``that car belongs to Alice'', while unlinkability ensures that no one can say ``the same car authenticated itself yesterday on highway 2''.  In this work, we refer to a PKI which guarantees anonymity and unlinkability as \emph{vehicular PKI} (VPKI).


Two other issues facing VPKIs relate to how it will operate in the field under potentially high mobility and intermittent connectivity to the infrastructure, and at large scales, for example, encompassing all vehicles in a large city or state.   Existing solutions such as SCMS incur large times for pseudonym certificate generation, which is necessary to refresh the credentials and support unlinkability.

\paragraph{Pseudonym Certificate Generation Time} Typically in VPKI systems, users authenticate themselves on the road using temporary credentials called \emph{pseudonym certificates} (PCs).  These PCs are periodically refreshed when the vehicle executes the PC generation protocol with a stationary road-side unit (RSU).  An extremely important (and often overlooked) feature in the design of VPKIs is the execution time of PC generation as this governs the possible use cases of the system.  Essentially, the issue is that a fast vehicle passing by an RSU would have only a very brief period of network connectivity (less than one second if traveling at highway speeds).  Therefore, if the PC generation protocol is too slow, the vehicle and RSU will fail to complete their protocol with high probability, and thus the vehicle would not reliably be able to update its PC.  This issue is aggravated by the fact that channel uncertainty in wireless networks intensifies in highly mobile environments~\cite{haenggi2009outage}.  Unreliability of PC generation at high speeds means, for example, that cars might wind up using the same PC for hours at a time, which harms system security.  Because of SCMS slow PC generation protocol, our results show that attempting to use SCMS in highway scenarios requires PCs to persist for 6.7 hours, while \name{} reduces this window to 18 minutes (a $\sim$22.5x privacy improvement).

\paragraph{Revocation}  Given the damage potential of a malicious user in VPKI, a revocation mechanism is needed to deactivate a malicious user's credentials, rendering the user unable to participate in the system.  Revocation solutions involving a global centralized certificate revocation list (CRL) are not ideal for VPKI.  On the one hand, the large scale and distributed nature of the system make it unreasonable to expect that all vehicles would constantly maintain an up-to-date local copy of the CRL.  On the other hand, network-based solutions where a vehicle queries a CRL on the cloud before interacting with another vehicle are not ideal because of latency and network intermittency.  Because of these issues, revocation is a major pain point of all prior VPKI systems.

\begin{figure}[t!]
\centering 
\includegraphics[width=.9\linewidth,keepaspectratio]{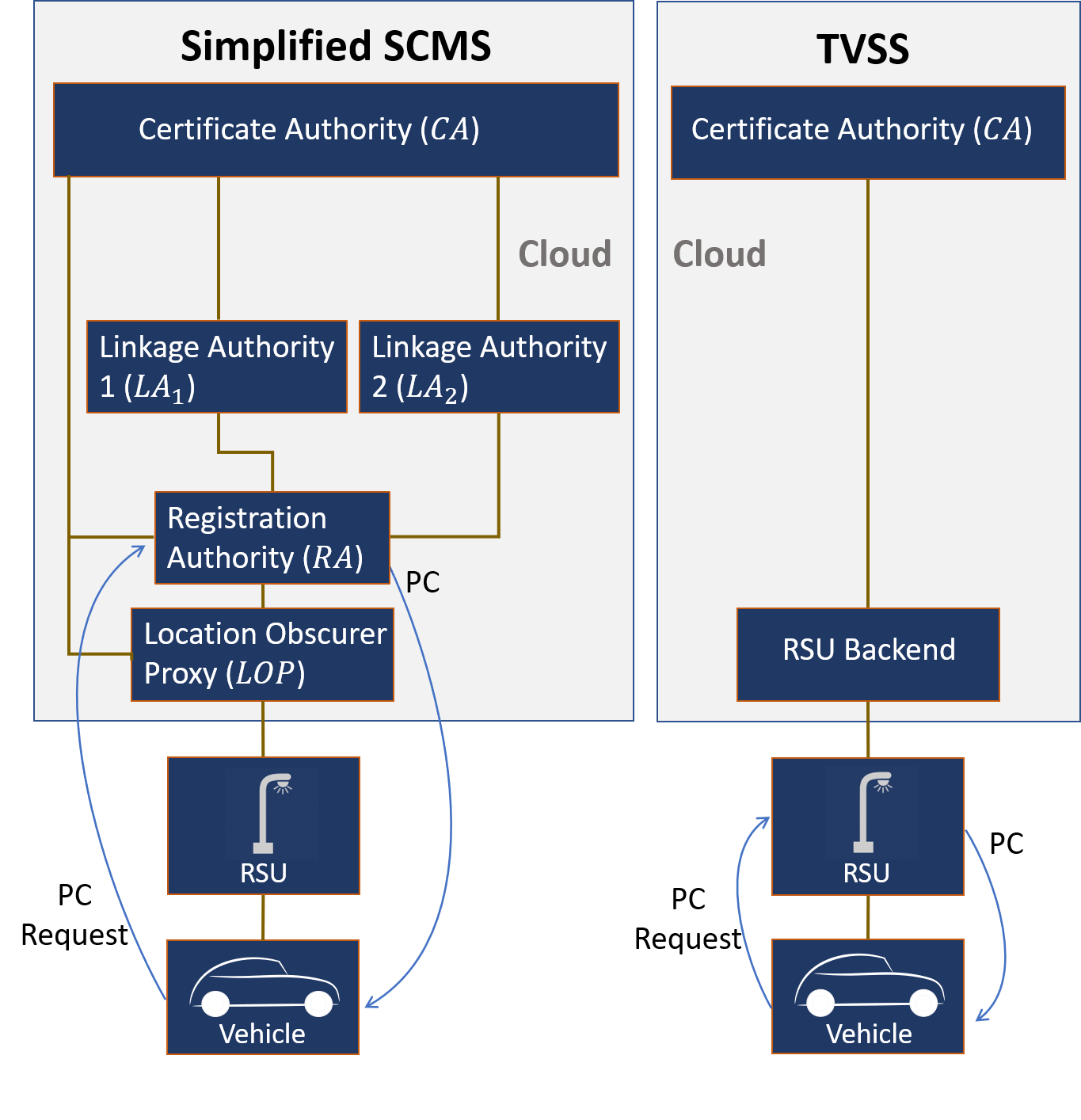}
\caption{Simplified SCMS and \name{} architectures.}
\label{fig:scms and dvss arch}
\end{figure}

\subsection{Our Contributions}
\label{sub:contribution}
\begin{itemize}[align=left]

\item[{\bf Edge-based VPKI:}] We propose Token-based Vehicular Security System (\name), a new system architecture for VPKI with properties which are essential for a large scale mobile PKI system.  The core novel feature of \namespace is that it takes advantage of the compute power of the network of roadside units (rather than using RSUs simply as a network of proxies connecting vehicles to the backend servers).  We find that computationally able RSUs fit seamlessly into VPKI, yielding improvements across the board. Specifically:


\begin{itemize}[align=left]
    \item[{\bf 1. Low latency PC generation:}] \namespace has a lightweight PC generation protocol consisting essentially of just a handshake between the vehicle and an RSU,
    as shown in Figure~\ref{fig:scms and dvss arch}.  In particular, PC generation requires \emph{no online involvement from the back-end}.  This enables new use cases which were unsupported by previous systems (\emph{e.g.}, high speed PC generation).
    
    \item[{\bf 2. Low Communication Revocation:}] Utilizing compute power of the RSUs enables reliable revocation where global CRLs are maintained by the (stationary) RSUs, while only short (location-specific) CRLs need to be shared with the vehicles.  The result is that revocation in \namespace requires drastically less total communication than in prior systems.

    \item[{\bf 3. Simple architecture:}] Passing computation to the RSUs greatly simplifies the system as a whole.  As shown in Figure~\ref{fig:scms and dvss arch}, the overall footprint of \namespace is much smaller than that of SCMS.  Thus, maintenance costs of \namespace would be much lower with no loss to security.
\end{itemize}

    \item[{\bf Formalizing VPKI Security:}] In order to foster future work on VPKI, we give a formal game-based security definition for VPKI incorporating unforgeability, anonymity and unlinkability.  Additionally, we consider other attacks on our system and show how to neutralize them.  This discussion includes a new type of attack called a \emph{clone attack} which was previously unconsidered in the VPKI literature, and which affects all prior systems.  Clone attacks are similar in spirit to \emph{sybil attacks}\footnote{Sybil attacks occur when a single vehicle obtains several different copies of valid credentials in order to pretend to be several different vehicles.}~\cite{douceur2002sybil}, and occur when an authorized vehicle shares its credentials with an unauthorized vehicle in an attempt for both cars to participate in the system in different locations.  Similar attacks have been considered in cryptocurrencies  ~\cite{chiesa2017decentralized, zhang2019double}, transportation toll collection~\cite{rupp2013p4r} and network authorization schemes~\cite{patel1997ticket}.  We show how to handle clone attacks against \name; no discussion or defense to clone attacks is given in other VPKI systems.
        
    \item[{\bf Open Source Testbed:}] We build and assemble a real testbed of on-board units (OBUs) and RSUs that have technical specifications similar to commercial OBUs and RSUs. Specifically, we set up the networking standard specifically designed for connected vehicles, IEEE 802.11p/dedicated short range communication.  Our OBUs and RSUs are open source and re-programmable and so hopefully will be useful to other research and application development.

    \item[{\bf VPKI Implementation and Field Experiments:}] We implement \name{} and other VPKI systems on our OBUs and RSUs and conduct a series of highway and in-city street experiments at different velocities ranging from 25mph to 85mph.  Our tests observed that \name{} achieves a 28.5x reduction in its PC generation latency and a 13x reduction in total communication during revocation compared to other systems. At extreme speeds, \name{} is 3.85x more likely to successfully complete PC generation and 6.5x more likely to successfully update its local CRL compared to other systems.

\end{itemize}

\section{System Overview}
\label{sec:vpki}
The goal of this section is to explain 1) how our system \namespace works at a high level; and 2) how the three features mentioned in Section~\ref{sub:contribution} are achieved.  We stress this is an oversimplified discussion, the full scheme is presented in the next section.

\vskip 2mm\noindent\textbf{System Players and the Parameter $T$.} The main players in \namespace are the vehicles, the distributed network of RSUs and the certificate authority (CA).  Additionally, we assume that the RSUs are all connected to a backend server called the ``RSU backend'' which communicates with, but is distinct from the CA (this is important for revocation).  We model the connection between the vehicles and the RSUs as a secure private channel (in reality this will be implemented using encryption).  Time in \namespace is broken into distinct periods of $T$ minutes each, for a system parameter $T$.  Roughly speaking, smaller $T$ means stronger security and greater overhead on the system.

\vskip 2mm\noindent\textbf{The Infrastructure/Backend Separation Assumption.} It is critical to security in \namespace to assume that the CA and the RSUs are controlled by separate entities which do not collude; we call this the \emph{infrastructure/backend separation assumption}.  Some version of this assumption is implicit in all prior work on VPKI.  Though not ideal, we believe the infrastructure/backend separation assumption necessary for security in \namespace is plausible since the CA would likely be controlled by a specialized security company while the RSUs would be part of the public infrastructure.  Designing a VPKI system which does not require any separation assumptions about infrastructure and the backend is an excellent open problem.

\vskip 2mm\noindent\textbf{System Setup.} Life begins for a vehicle when it obtains an enrollment certificate EC from the CA.  We think of this as occurring at the manufacturing plant, once during the vehicle's lifetime.  

\vskip 2mm\noindent\textbf{Token Generation.} Once a vehicle possesses its enrollment certificate, it can request a batch of tokens from the CA which it will later trade in to the RSUs in order to get PCs.  In order to request tokens, the vehicle simply authenticates itself using the EC in order to receive a large number of tokens (say 2,880 for a one month's supply when $T=15$ minutes).  Each token is signed using CA's private signing key and can be validated by verifying the signature using CA's public signing key.  Each token is valid for one specified period of $T$ minutes; when the final token expires, $v$ will need to request new tokens from CA.  Token generation can be performed offline.


\vskip 2mm\noindent\textbf{PC Provisioning.} Once a vehicle has tokens, it can request PCs from any RSU.  To do this, the vehicle simply presents the token for the current time period to the RSU who verifies authenticity against CA's public credentials and, if valid, returns a PC.  The PC is signed by the RSU and marked with a geographical tag corresponding to the RSU's location, and is valid only when nearby this location and during the same time window that the token is valid for.  This means that new PCs must be requested every $T$ minutes.  The geographic radius of validity should be an upper bound on the distance one can drive in $T$ minutes.  The RSU sends the token it received to the RSU backend to check for duplicates.  If the same token is used twice to generate two different PCs in two different geographical areas (\emph{i.e.}, if a clone attack is being launched), then the RSU backend will notice and will begin the revocation procedure for this vehicle (described next).

Notice that from the vehicle's point of view, PC generation consists of a simple ``handshake-type'' interaction with the RSU.  This is in stark contrast to SCMS where PC generation requires the vehicle's messages to be forwarded all the way to the RSU backend which is four ``network hops'' away from the vehicle.  The ability to perform PC generation over a short-term connection unlocks use cases which are not supported by SCMS.  For example, our experiments in Section~\ref{sec:evaluation} demonstrate that PC generation in \namespace can be performed at high speeds on the freeway, while this would not be possible in SCMS.  Indeed, our experiments indicate that in these driving conditions, our PC generation protocol is at least $10x$ less likely to fail than PC generation in SCMS.  This is an important improvement since resolving the issues resulting from failed PC generation attempts consumes extra system resources.  Additionally, because PC generation can be performed with a much weaker connection between the vehicle and RSU, the frequency with which a connection occurs which can support PC generation increases considerably.  This allows us to set our system up so that PCs are refreshed more frequently (every fifteen minutes, rather than once per week) which translates to better security and easier revocation.

\vskip 2mm\noindent\textbf{Revocation.} When the PC of a malicious vehicle is identified, the RSU backend and the CA are alerted, and they cooperate to recover the current and future tokens of the offending vehicle.  These tokens are shared with the RSUs who update their token blacklists ($\TRL$s), thus preventing the offending vehicle from obtaining any new PCs in the future.  In order to deactivate the current PC, the CA shares the offending PC \emph{only with the RSUs which are in the geographic radius of the misbehaving vehicle}, these RSUs in turn share the PC with all vehicles in their region, who update their local CRL (or what we also refer to as pseudonym certificate revocation lists, PCRLs) and make sure to avoid the offending vehicle.

Note that the honest vehicles receive only the PCs of the misbehaving cars in their geographical area, not the list of all misbehaving vehicles in the whole system.  Moreover, this list must only be maintained for the remainder of the time period, after which point it can be discarded.  Thus, the amount of revocation information which needs to be downloaded by each vehicle is small.  The large $\TRL$s which must be maintained by the RSUs (consisting of all current and future tokens of all offending vehicles in the system) represent less of a problem as the RSUs are stationary and so should have a stable connection.  Our experiments in Section~\ref{sec:evaluation} demonstrate that this change to requiring the vehicles only to maintain a geographically-based CRLs leads to a system-wide $13x$ savings in total communication size.

We remark that while the idea to use geographic PCs to improve revocation seems, at first glance, to be a generic solution which can be applied in any VPKI, this is not the case.  The key feature of \namespace which makes it possible is the short lifespan of the PCs (\emph{i.e.}, small $T$). In SCMS, the PCs are live for an entire week and so it is not possible to constrain a PC to a small geographic region (one week is enough time to drive from Lisbon, Portugal to Vladivostok, Russia).  The short lifespan of PCs is only possible in \namespace because of the system-wide efficiency improvements gained by passing computation to the RSUs.

\vskip 2mm\noindent\textbf{Security.} Anonymity and unlinkability both hold in \namespace intuitively because the tokens which are used to get the PCs have no information about the vehicle.  For example, even if the RSU is corrupt and can connect the PC to the token, there is no way to connect the token to the vehicle's enrollment certificate as long as the RSU does not collude with the CA.  Thus, anonymity should hold under the infrastructure/backend separation assumption; unlinkability should hold for similar reasons.  Formal security definitions and proofs are given in Section~\ref{sec:security}.

\section{\name\ System}
\label{sec:new system}
In this section, we start by describing the system model and assumptions. We then illustrate the detailed protocols forming our system, namely ($\setup$, $\tokengen$, $\pseudogen$, $\revoke$). We finally discuss some implementation considerations.

\begin{figure}[t]
\centering 
\includegraphics[width=\linewidth,keepaspectratio]{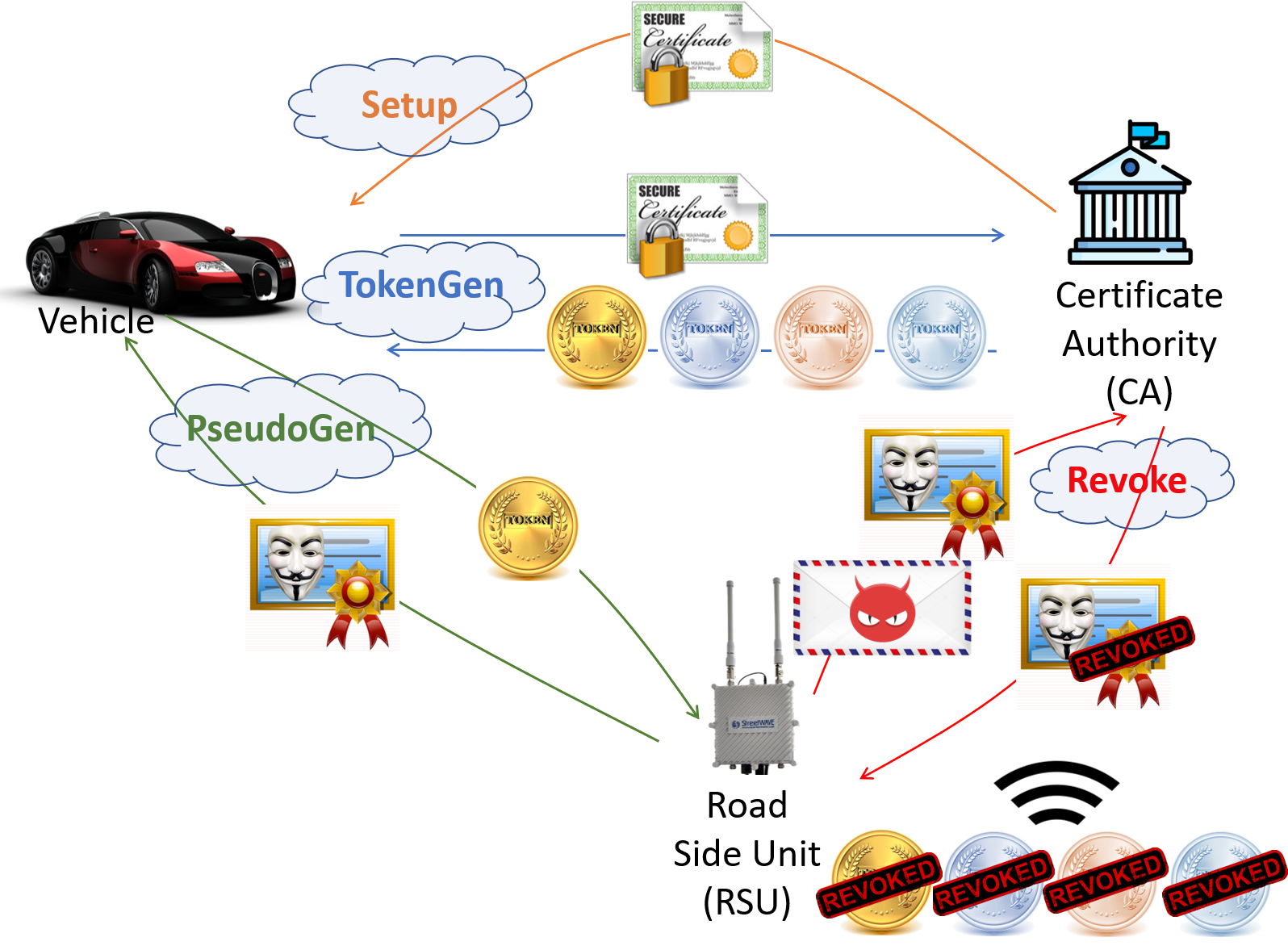}
\caption{Overview of all protocols of \name{}.}
\label{fig:DVSS all phases overview}
\end{figure}

\subsection{System Model and Assumptions}
As mentioned before, the \name{} system consists of a network of RSUs and backend servers residing in the cloud. The backend servers are the RSU backend and the certificate authority (CA).  The CA is the root of trust in the VPKI and it is responsible for providing vehicles with $\ltc$s and tokens while providing RSUs with signing certificates. Vehicles use $\ltc$s to obtain tokens from the CA while RSUs use their signing certificates to generate and sign PCs for vehicles in exchange for tokens (see Figure \ref{fig:DVSS all phases overview}).  Vehicles use dedicated short range communication (DSRC/IEEE 802.11p) to communicate with one another, and they can reach the internet and the backend servers through the RSUs only.  The communication channels between system's entities (i.e. vehicles, RSUs, backend servers, etc) are assumed to be always secure (e.g., over TLS). The system protocols described next make use of a standard digital signature scheme.

\subsection{System Protocols}
\label{sub: System Protocols}
The four protocols $(\setup,\tokengen,\pseudogen,\revoke)$ make up the \namespace system.

\begin{enumerate}[align=left]
    \item[$\bullet$ $\setup${\bf :}] This allows a vehicle $v\in\V$ to acquire a long-term enrollment certificate, $\ltc_v$, which acts as the vehicle identity.  Explicitly $\ltc_v$ is a signing key pair consisting of a private signing key and a public verification key, and additionally a signature on the public verification key using the CA's private signing key.  The vehicle $v$ will use $\ltc_v$ to request services from the $\CA$ when needed using a TLS-style handshake. The enrollment certificate $\ltc_v$ has a long expiration time and $v$ probably gets it once in its lifetime or upon ownership transfer.  Figure~\ref{fig:DVSS all phases overview} shows an overview of this protocol, the full protocol is given in Algorithm~\ref{alg:setup}.

    \item[$\bullet$ $\tokengen${\bf :}] This provides a vehicle  with a list of authorization tokens $\T = \{\tau_1,\cdots,\tau_N\}$ that can be used to  anonymously request pseudonym certificates from the RSUs.  Each token $\tau\in\T$ is active during a specified time window only, and so once the final token in $\T$ expires, the vehicle will have to rerun this procedure to get more.  Explicitly, each token is a random nonce, a time window, and a signature on the nonce/time window pair.  The vehicle obtains the tokens after completing a TLS-style handshake with the CA using $\ltc_v$.  Figure~\ref{fig:DVSS all phases overview} shows an overview of $\tokengen$, and Algorithm~\ref{alg:tokenGen} shows the full protocol.

    \item[$\bullet$ $\pseudogen${\bf :}] This is used by vehicles to request PCs from the RSUs in exchange for tokens.  Explicitly, a PC is a signing key pair consisting of a private signing key (generated by the vehicle and kept secret), and a public verification key, and a signature on the public verification key using the RSU's secret signing key.  The vehicle simply presents its token and its verification key to the RSU who validates the token and then signs the verification key and returns the PC. The RSU also shares this token and the certificate with the RSU backend to detect if token double use occurs; this allows our system to detect and throttle clone attacks.  Figure~\ref{fig:DVSS all phases overview} shows an overview of $\pseudogen$, the full protocol is shown in Algorithm~\ref{alg:pseudogen}.

    \item[$\bullet$ $\revoke${\bf :}] When a vehicle is identified as adversarial, the RSU backend and the CA cooperate to run this protocol to deactivate the offending vehicle's credentials.  This works, as described in the previous section, by updating the RSU $\TRL$s to include all future tokens of the offending vehicle, by updating the vehicle CRLs of all vehicles which are nearby the offending vehicle to include the malicious PC, and by updating the CA blacklist $\BL$ to include the enrollment certificate ($\ltc_v$).  Figure~\ref{fig:DVSS all phases overview} shows an overview of $\revoke$; the full protocol is in Algorithm~\ref{alg:revoke}.
\end{enumerate}

\begin{algorithm}[t]
\begin{algorithmic}[1]
\State Compute $(\vk_v, \sk_v)\leftarrow\keygen()$;
\State Compute $\sigma_v \gets \sign_{\sk_\CA}(\vk_v)$
\State $\ltc_v \gets (\vk_v, \sigma_v)$ 
\State\textbf{Output} $\ltc_v$
\end{algorithmic}
 \caption{$\setup(\sk_\CA)$}
 \label{alg:setup}
\end{algorithm}

\begin{algorithm}[t]
\begin{algorithmic}[1]
\State Parse $\ltc_v =(\vk_v, \sigma_v)$;
\State Parse $Req =(\st', \et')$;
\If{$\revoked(\ltc_v)=\text{False} $ \; \& \; $\hastokens(\ltc_v, \st',\et')=\text{False}$}
\State Compute $\T = \{\tau_1, \cdots, \tau_N\}, \text{\;where\;} \tau_j\leftarrow(\id_j, \tw_j, \sigma_j)$
\State $tc \gets (\id_j, \tw_j)$
\State $\sigma_j \gets \sign_{\sk_\CA}(tc)$
\State\textbf{Output} $\T$
\Else
\State\textbf{Output} $\perp$
\EndIf
\end{algorithmic}
 \caption{$\tokengen(\ltc_v, Req, \sk_\CA)$}
 \label{alg:tokenGen}
\end{algorithm}

\begin{algorithm}[t]
\begin{algorithmic}[1]
\State Parse $\tau=(\id, \tw, \sigma)$
\State Compute $(\vk, \sk)\stackrel{R}{\leftarrow}\keygen() $
\If{$\timevalid(\tau)=\text{True} \; \& \; \revoked(\tau)=\text{False}\; \& \; \verify_{\vk_\CA}((\id, \tw), \sigma)=\text{True}$}
\State Compute $\sigma' \gets \sign_{\sk_{RSU}}(\vk)$
\State Compute PC$ \gets (\vk,\sigma')$
\State\textbf{Output} PC
\Else 
\State\textbf{Output} $\perp$
\EndIf

\end{algorithmic}
 \caption{$\pseudogen(\tau, \sk_{RSU}, \vk_\CA)$}
 \label{alg:pseudogen}
\end{algorithm}

\begin{algorithm}[t]
\begin{algorithmic}[1]
\If{$\deciderevoke(\text{PC})=\text{True}$}
\State $\tau_j\leftarrow\gettoken(\text{PC})$
\State $\ltc_v\leftarrow\findltc(\tau_j)$
\State $\update(\BL, \ltc_v)$
\State $\update(\TRL, \{\tau_j, \cdots, \tau_N\})$
\State $\update(\text{PCRL}, \text{PC})$
\State \textbf{Output} $(\BL, \TRL, \text{PCRL})$
\EndIf
\end{algorithmic}
 \caption{$\revoke(\text{PC})$}
 \label{alg:revoke}
\end{algorithm}


\subsection{Implementation Considerations}

\vskip 2mm\noindent\textbf{RSU Blackout Areas.} Suppose a vehicle needs to travel to an area with limited RSU deployment for an extended period, but wants to precompute PCs for use during its trip.  Normally the RSUs in our system will only give a PC for the current time window.  However, in such a situation, our system could allow an RSU to grant a batch of PCs in exchange for a batch of tokens and the vehicle would be able to authenticate itself as usual.  Note however that the unlinkability guarantee would fail to hold as the RSU would likely infer that the batch of PCs all belong to the same vehicle.  


\vskip 2mm\noindent\textbf{Linked Tokens via Hash Chains.} 
We propose a novel approach that allows for generating token ids for a single vehicle $v$ that look random and independent   but can still be selectively linked using a secret that is  kept at CA only. Particularly, it provides two advantages: 1) It allows CA to publish a single $\tau$ id (and a secret) per revoked $v$, which can significantly reduce token blacklist (\TRL) bandwidth requirements since the $\TRL$ size becomes based on the number of revoked vehicles rather than revoked tokens. 2) The CA can keep a single $\tau$ id (and a secret) at any time window and derive the ids of tokens of future time intervals; the same approach can also be utilized by an RSU when tokens are revoked. This can significantly reduce the storage requirements of keeping such token ids at CA and RSU. 

\begin{figure}[t]
\centering
\includegraphics[width=\linewidth,keepaspectratio]{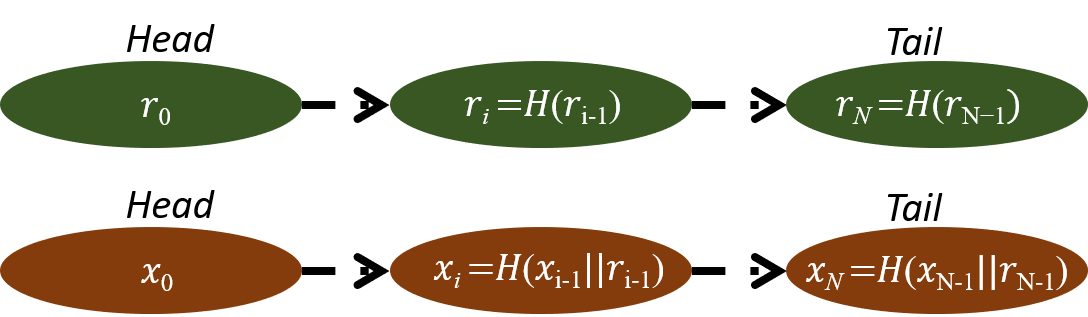}
\caption{Our new hash chain mechanism.}
\label{fig:New Hash Chain}
\end{figure}

Before explaining our hash chain approach (Figure~\ref{fig:New Hash Chain}), we first discuss the intuition behind it.  The approach utilizes hash chains because each hash value of theirs look random and thus can be used as a $\tau$ id. However, since the hash function $H$ is public, $\rsu$ can determine if two hashes $x_i, x_{j}$ belong to the same chain by using one to derive the other using $H$, which violates unlinkability. To fix this issue,  CA uses a per vehicle secret $r$ as part of the $H$ input  to traverse the $\tau$ id hash chain of $v$; thus, rsu is unable to link $\tau$ ids without knowledge of $r$ due to the inversion hardness assumption of $H$. Additionally,  upon revocation and revealing the secret $r$,
our mechanism needs to preserve unlinkability of $v$ before it was revoked (e.g., protect privacy of a stolen $v$). To achieve this property, the per vehicle secret $r$ is also constructed as another hash chain as shown in Figure~\ref{fig:New Hash Chain}; thus, when $v$ is revoked at a time interval, only corresponding hash values from both hash chains are revealed,  which only allow rsu to derive current and future revoked tokens of $v$. 

Our hash chain mechanism works as follows; At vehicle $\setup$,  CA picks two random values, $x_0$,$r_0$, called heads. Upon executing $\pseudogen$, CA uses $H$ and the heads $x_0$,$r_0$ to create two hash chains of length $N+1$, where $N$ is the number of tokens generated as a result of this request; the next $\pseudogen$ request heads become $x_N$,$r_N$. In the hash chain of secrets (i.e., $r_0,\cdots,r_N$), at the $i^{th}$ position, the value is calculated by $H(r_{i-1})$. The second hash chain, which produces the $\tau$ ids, makes use of the values from the first chain. Specifically, the value at the $i^{th}$ position is calculated by applying $H(x_{i-1} || r_{i-1})$.  Whenever a malicious vehicle is detected, then the remaining tokens are revoked by publishing the corresponding input values to the current token $x_i$ being revoked, that is, $(x_{i-1}, r_{i-1})$. With this, the rsu will be able to invalidate the other remaining tokens acquired by the revoked $v$, which allows rsu to detect and refuse providing new PCs to revoked vehicles. 

\section{Security}
\label{sec:security}
In this section we discuss the security of our new system \name.  We begin in Section~\ref{sec:sybilclone} by considering Sybil and clone attacks.  Then in Section~\ref{sec:securitygame} we give a formal game-based security definition for VPKI which captures anonymity and unlinkability.  Then, in Appendix~\ref{sec:secproof}, we prove that \namespace satisfies the security definition.

\subsection{Sybil and Clone Attacks}
\label{sec:sybilclone}

\noindent\textbf{Sybil Attacks.} A Sybil attack occurs when a single malicious vehicle impersonates many independent vehicles.  Sybil attacks are a major concern in SCMS as the vehicles are given many valid PCs simultaneously and instructed, but not forced, to use them one at a time.  However, in our system vehicles get only one valid PC at a time and so a Sybil attack is only possible if the malicious vehicle obtains (either consensually or illicitly) active PCs from other actual users.  Thus, launching a Sybil attack on \namespace is much more difficult than on SCMS since it requires either convincing several other users to misbehave, or it requires obtaining their credentials via some other means.  In this case, if the Sybil attack were noticed, the offending vehicle's credentials would be revoked, along with the credentials of all vehicles whose PCs were used.

\vskip 2mm\noindent\textbf{Clone Attacks.} A clone attack occurs when a malicious vehicle shares its token with another vehicle in another geographic area and both of them use the token to obtain a valid PC and participate in the system.  Clone attacks are more problematic for our system since there is nothing tying the token to the vehicle, and so there is no way for the second RSU to notice that the token does not ``belong'' to the clone vehicle.  However, when the RSUs send the tokens to the RSU backend for analysis, the RSU backend will discover the clone attack since it will notice that the same token was used in two different places.  Immediate revocation will follow, and so the next time the offending vehicle tries to get a PC from a RSU, it will find that its tokens no longer work.


\subsection{VPKI Security}
\label{sec:securitygame}
In this section, we formalize the security notions of anonymity and unlinkability for a VPKI system using a game-based security definition.

\vskip 2mm\noindent\textbf{Security Game Overview.} Let $(\setup, \tokengen, \pseudogen)$ be the algorithms from a TVSS system (the $\revoke$ procedure plays no part in this game), and furthermore let $({\sf Sign},{\sf Verify})$ be the sign and verify algorithms for the signature scheme used in the VPKI (recall that the PCs consist of a signature key pair and a signature on the public key using the RSU's secret key).  In the security game, we give the adversary $\mathcal{A}$ full control of all parties in the system: $\mathcal{A}$ can create vehicles using $\setup$, $\mathcal{A}$ can request tokens/PCs as $v$ using $\tokengen$/$\pseudogen$ which $\mathcal{A}$ can choose to either expose or not.  Eventually $\mathcal{A}$ decides how it will attempt to win the game; it has three options, it can win via forgery, it can break anonymity, or it can break unlinkability.  The challenger $\mathcal{C}$ then issues a corresponding challenge to $\mathcal{A}$, and $\mathcal{A}$ attempts to win the game.  For example, if $\mathcal{A}$ decides it wants to win via anonymity, then $\mathcal{A}$ will specify two different vehicles and $\mathcal{C}$ will a PC of one of them and $\mathcal{A}$ must guess to which vehicle the PC belongs.  The security game is shown in Figure~\ref{fig:securitydef}, the full set of legal queries and winning conditions are shown in Figures~\ref{fig:initialization} and~\ref{fig:winning}.  Definition~\ref{def:sec} simply says that the VPKI scheme is secure if no efficient adversary can win with any non-negligible advantage.  Note that $\mathcal{A}$ can trivially win the anonymity or unlinkability version of the game by guessing randomly, therefore advantage in this case means $\prob{}\bigl[\mathcal{A} \text{ wins}\bigr]-1/2$, while in the forgery version of the game $\mathcal{A}$'s advantage is simply $\prob{}\bigl[\mathcal{A}\text{ wins}\bigr]$. 



\threenmcbox

\winning

\begin{defn}
\label{def:sec}
We say that a VPKI scheme is \emph{secure}, if for all efficient adversaries $\mathcal{A}$, $\mathcal{A}$'s advantage in winning the VPKI security game is negligible.
\end{defn}

A thorough proof that our proposed scheme, \namespace, is secure is provided in  Appendix~\ref{sec:secproof}. We provide a security reduction for all the three challenges: forgery, anonymity, and unlinkability.

\initialization

\section{Experimental Testbed}

\begin{figure}[t]
\centering
\begin{subfigure}{.20\textwidth}
\includegraphics[width=\linewidth,keepaspectratio]{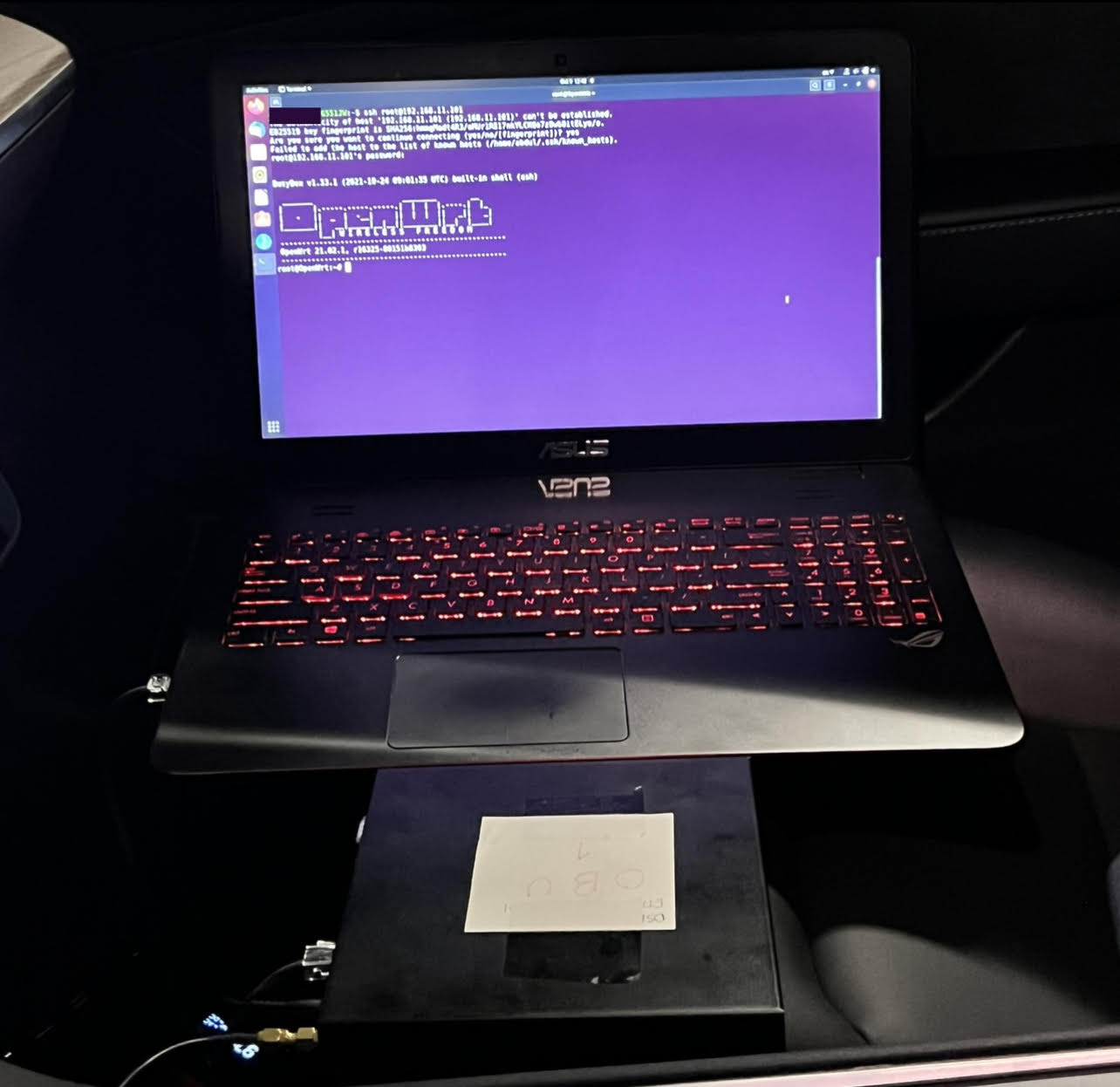}
\caption{}
\end{subfigure}
\begin{subfigure}{.27\textwidth}
\includegraphics[width=\linewidth,keepaspectratio]{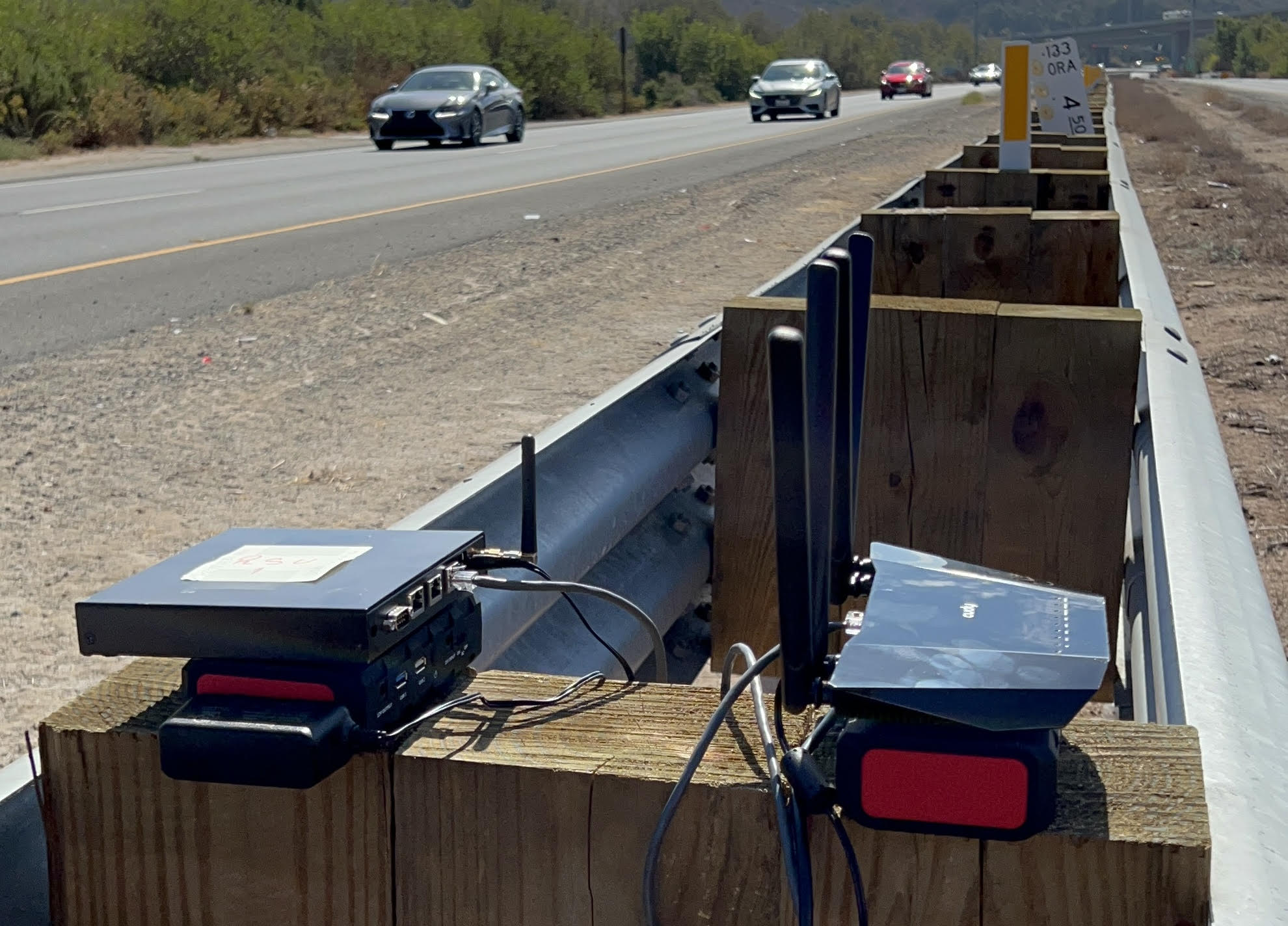}
\caption{}
\end{subfigure}
\caption{Highway testbed: a) An OBU controlled by a laptop, b) An RSU connected to a cellular router on the highway.}
\label{fig:testbed}
\end{figure}

In this section, we first present the hardware used for our experiments. We also provide technical details on our connected vehicles networking technology IEEE 802.11p/DSRC setup. We then discuss implementation details and issues when conducting experiments on connected vehicles. We finally explain the design of our field experiments on connected vehicles; Figure~\ref{fig:testbed} shows a snapshot of our highway testbed.

\subsection{Hardware}
\label{sub:Hardware}
We obtain 5 PC Engine APU1D4 embedded devices that can act as either on-board units (OBUs) or road-side units (RSUs). APU1D4 devices are carefully chosen as they have technical specifications that are similar to commercial OBUs and RSUs shipped by well-known vendors like commsignia (ITS-OB4 OBU and ITS-RS4 RSU) and Savari (MW-1000 OBU and SW-1000 RSU). Each of our PC Engine APU1D4 devices has AMD G series T40E - 1 GHz dual Bobcat core, 4 GB DDR3-1066 DRAM and a 16GB SSD hard drive. We build APU1D4 based OBUs and RSUs from scratch so that we can have an open source environment where we can implement various applications/VPKI protocols to get real field measurements of their performance in a vehicular environment. Our OBUs and RSUs run on a patched version of OpenWrt 21.02.1, a well-known firmware mainly used with embedded and networking devices and is utilized by Savari commercial OBUs and RSUs.

Moreover, we set up two cloud servers at a well-known cloud provider, Digital Ocean, acting as the certificate authority (CA) and registration authority (RA), which are needed by alternative cloud-based VPKI systems. Each of the CA and RA runs on an independent server with 4 CPUs, 8GB of DDR3 RAM and 160GB of SSD hard drive.   We also obtain Cudy N300 routers, which are equipped with LTE SIM cards from a cellular service provider; RSUs are connected to cellular routers to enable internet connectivity. Whenever OBUs need to reach the internet (e.g., for CA, RA ...etc), they make use of RSUs to relay their requests to the cellular router.

\subsection{IEEE 802.11p/DSRC Setup}
In order to enable our OBUs and RSUs to use the networking standard IEEE 802.11p (DSRC) - which is specifically designed for automotive communication, we install UNEX DHXA-222 wireless network interface cards (NICs), which are compatible with IEEE 802.11p/DSRC. These NICs have been validated and proven to deliver reliable IEEE 802.11p/DSRC communication~\cite{raviglione2019characterization, raviglione2021experimental}. To have IEEE 802.11p/DSRC properly set up, we configure our NICs to use the Outside Context of a BSS (OCB) mode; this allows direct communication between OBUs and RSUs (i.e., V2I/V2V) in an instant fashion with no association/authentication handshakes needed at the link layer. We also use the vehicular standard 5.9GHz frequency band with a channel width of 10MHz (namely, channel 178 at 5.890GHz) as specified by the standard. While the NICs allow us to set bit rates of 3-27Mbps, we configure our testbed to have the baseline standard DSRC bit rate of 6Mbps~\cite{adams2016development}. We also use the IEEE 802.11p compatible antennas from MobileMark (for OBUs) and PulseLarsen (for RSUs) and set the transmission power to 15dBm.

\subsection{Implementation}
We develop Bash scripts that utilize the ICMP protocol in order to  determine the contact time between OBUs and RSUs when CVs pass by RSUs. We also use Python to implement 3 VPKI systems to measure their performance, namely 1) \name{}, 2) SECMACE~\cite{khodaei2018secmace}, a recent VPKI system, and 3) SCMS~\cite{Brecht2018SCMS}, the US VPKI standard.

In SCMS, an OBU can request a fresh PC only from the cloud. In SECMACE, the OBU requests tickets from the cloud. Then the OBU uses these tickets to obtain fresh PCs from a Pseudonym Certificate Authority (PCA). Note that in SECMACE, the authors assumed the PCA to be in the cloud as well, but in order to perform a fair comparison between SECMACE and \name{}, we decided to place the implementation of PCA for SECMACE inside the RSUs. In \name{}, the OBU requests tokens from the cloud and uses these tokens to obtain fresh PCs from the RSUs. The PC in both SECMACE and \name{} is valid only within a specific region but the main difference is that when a vehicle leaves the validity region in SECMACE, the OBU must request new tickets from the cloud, while in \name{} the OBU simply uses a token to request a new PC from the RSU for the new region.


In all of the cloud based VPKIs, any interaction with the cloud must go through an RSU. Each RSU is equipped with a switching and a routing fabric and configured with a routing table to quickly and efficiently forward any OBU request to a cloud-based IP address via the connected 4G gateway (see Section \ref{sub:Hardware} and Figure \ref{fig:testbed}). Note that from an OBU point-of-view, the RSU is the gateway to the cloud servers. In addition, all communications between OBUs, RSUs and cloud servers use reliable and secure channels TCP/TLS1.3, and for that the mainstream Python libraries socket and ssl are utilized. We use hashlib library for hashing using SHA256. 

For public key cryptography, we utilize PyNaCl, a Python binding to libsodium, which is a fork of the Networking and Cryptography library (NaCl). We use the library to implement public key encryption using the Elliptic Curve Integrated Encryption Scheme (ECIES) algorithm with the curve25519 elliptic curve and EdDSA digital signature algorithm with the ed25519 elliptic curve. Both algorithms use 256-bit long private keys and are adopted by SCMS. All VPKI components (OBU, RSU, CA and RA) use certificates for authentication.


The nature of experimenting on fast moving vehicles makes the contact time between OBUs and RSUs (i.e., gateways) limited as will be shown in Section~\ref{sec:evaluation}. We initially experienced difficulties measuring cloud-based VPKI systems accurately since they tend to have  longer round trip time (RTTs) and OBUs could try to communicate with cloud by sending a packet just right before it is in contact with the RSU and wait till after it is out of coverage. We address this issue by letting OBUs periodically create a new independent thread and send a new request to the cloud servers. We set the timeout to 30ms for OBU-RSU communication and 2s for OBU-cloud communication; a new request thread is created every 500ms to make sure VPKIs get a chance to make a request while in contact with the RSU and to not saturate the receivers.

\subsection{Experimental Scenarios}
\label{sec:Experimental Scenarios}
We evaluate all 3 VPKI systems under the following scenarios: 1) highway and 2) in-city scenarios. For the highway scenario, the highway has 4 lanes (2 are inbound and the other 2 are outbound) and a median strip in the middle. 1 OBU is installed in a vehicle and 4 RSUs are placed on the median strip (i.e., middle of highway). The vehicle passes by the 4 geographically distributed RSUs at various highway speeds ranging from 55mph to 85mph. For the in-city scenario, the street has 2 opposing lanes with no median strip; 1 OBU is installed in the vehicle and 4 RSUs are placed on the side of the street. The vehicle passes by the 4 geographically distributed RSUs at relatively slow speeds, namely 25mph and 35mph. In both scenarios, RSUs do not have overlapping coverage and are connected to cellular routers, and the OBU antenna is placed outside the vehicle for better signal transmission and reception. In each scenario, we conduct two sets of experiments, PC Procurement (obtaining a PC from the VPKI) and PCRL Download (retrieve a complete and up-to-date PCRL from the RSU). For each experiment and each speed, we run and repeat the experiment 25 times which generate 100 results per run (25 trials x 4 RSUs).
\section{Evaluation}
\label{sec:evaluation}
In this section, we present a real field-based study of \name\ and its two alternatives (SECMACE and SCMS) using the measurements collected by our testbed of OBUs, RSUs, and cloud servers on highway and in-city streets at a range of driving velocities as mentioned in Section~\ref{sec:Experimental Scenarios}. As described in Section \ref{sec:new system}, \name\ is edge-based, pushing primary services to the edge of the network (the RSU) to enable operation compatible with potentially short connection times. As a result, we avoid communicating with centralized back-end services during critical operations, shortening response time and increasing the reliability of the offered services. A primary advantage of \name\ in a V2I communication setting, is that it enables anonymized reauthentication of connected vehicles (CVs) as they drive by RSUs. A CV may have access to the infrastructure services for a short period of time (depending on its speed, range, and location) before it exists the coverage of the serving RSU. The purpose of our experiments is to measure the effect of the network coverage time on the service performance and achieved security for different traveling speeds. We additionally discuss the scalability of actual RSUs if \name{} were to be deployed. For all VPKIs, we collected 100 field samples for each of the experiments in order to get statistically reasonable results. Next we present our evaluation campaigns; each campaign evaluates a specific metric in the design of the VPKIs.

\begin{table}[t]
\centering
    \begin{tabular}{lSSSS}
    \toprule
    \multirow{1}{*}{Speed} &
      \multicolumn{4}{c}{\makecell{Coverage Time (seconds)}} \\ & {Minimum}
    &  {Median} & {Average} & {Maximum}        \\
      \midrule
        85mph & <0.1 & 1 & 1.39 & 5   \\
        75mph  & 0.32  & 1 & 1.91 & 5  \\
        65mph & 0.61 & 2 &  2.07 & 6 \\
        55mph & 1 & 3 & 3.30 & 8 \\
        35mph & 3 & 6 & 5.75 & 10 \\
        25mph & 3 & 6 & 6.30 & 12   \\
    \bottomrule
  \end{tabular}
   \caption{Highway and in-city OBU-RSU coverage time.}
    \label{tab:Coverage time}
    \end{table}

\vskip 3mm
\textbf{Evaluation 1: Measuring Basic Coverage Times.} In the first evaluation, we aim to capture the time a vehicle stays in the coverage of an RSU under different speeds and directions to study the implications of vehicle network connectivity time on VPKI. The results are presented in Table \ref{tab:Coverage time}. As we can see, in a worst-case scenario, for a vehicle traveling at a high speed (85mph) that passes by an RSU from the outer region of its coverage area (i.e. far away from the RSU), the coverage time can be lower than 100ms. 
Our results also show that vehicles driving on the highway (55mph-85mph) have coverage time with the RSU ranging from 1.39s to 3.30s, which is very limited. This suggests that all communications with the RSU need to be completed in a very short period before vehicles are out of coverage. Furthermore, the operations that require the backend/cloud involvement require the link between the RSU and backend servers to always be reliable, which might be infeasible in vehicular environments (e.g., RSU might have to rely on an unstable wireless service, with only intermittent connectivity.) In addition, Table~\ref{tab:Coverage time} measures the coverage times for vehicles driving inside the city at relatively slow speeds (25mph-35mph). While vehicles have more contact time with RSUs (5.75s to 6.30s) in the city, the results show that there are cases where contact time can be limited (as low as 3s). This is an expected outcome in realistic testbeds as the conditions of wireless channels fluctuate in dynamic environments such as vehicular networks.

\begin{figure}[t]
\centering
\includegraphics[width=.8\linewidth,keepaspectratio]{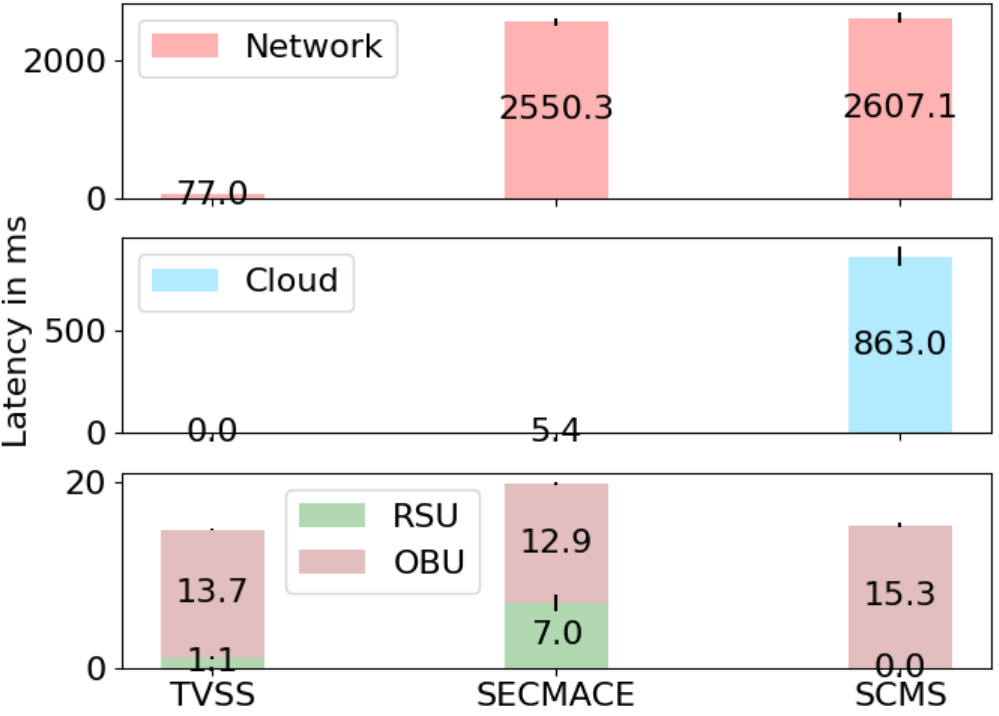}
\caption{PC generation latency of VPKI systems.}
\label{fig:Pc Gen Latency}
\end{figure}

\vskip 3mm
\textbf{Evaluation 2: PC Generation Latency.} 
In order to get a better understanding of the design implications of VPKI systems, we measure the end-to-end latency of a PC request operation using different VPKI designs; \name{}, our proposed VPKI system, SECMACE, a recent VPKI system, and SCMS, the US VPKI standard. Recall that for PC generation, \name{} completely relies on the RSU while SECMACE relies on both the RSU and a cloud-based CA. For SCMS, it relies completely on the cloud servers. Figure~\ref{fig:Pc Gen Latency} presents the elapsed time at the OBU, RSU, network and cloud servers for completing a single PC generation request. For all VPKI protocols, the most dominating factor is the network time. Our results show that relying completely on the edge for refreshing PCs, as in \name{}, helps cut down network latency by 33.5x compared to relying on the cloud.  Thus, these results show that \name{} reduces total PC generation latency by 28.5x and 38.5x compared to SECMACE and SCMS, respectively. The figure additionally indicates that RSU overhead is reduced by 6.4x in \name{} compared to SECMACE; this has bad implications on the scalability of RSUs for SECMACE as will be discussed later in this section. Even though SECMACE PC generation cannot rely completely on the edge for security reasons, \name{} cuts down the combined edge overhead (OBU and RSU) by 1.36x compared to SECMACE. This implies that \name{} is a more efficient solution compared to its alternatives.

\begin{figure}[t]
\centering
\includegraphics[width=.8\linewidth,keepaspectratio]{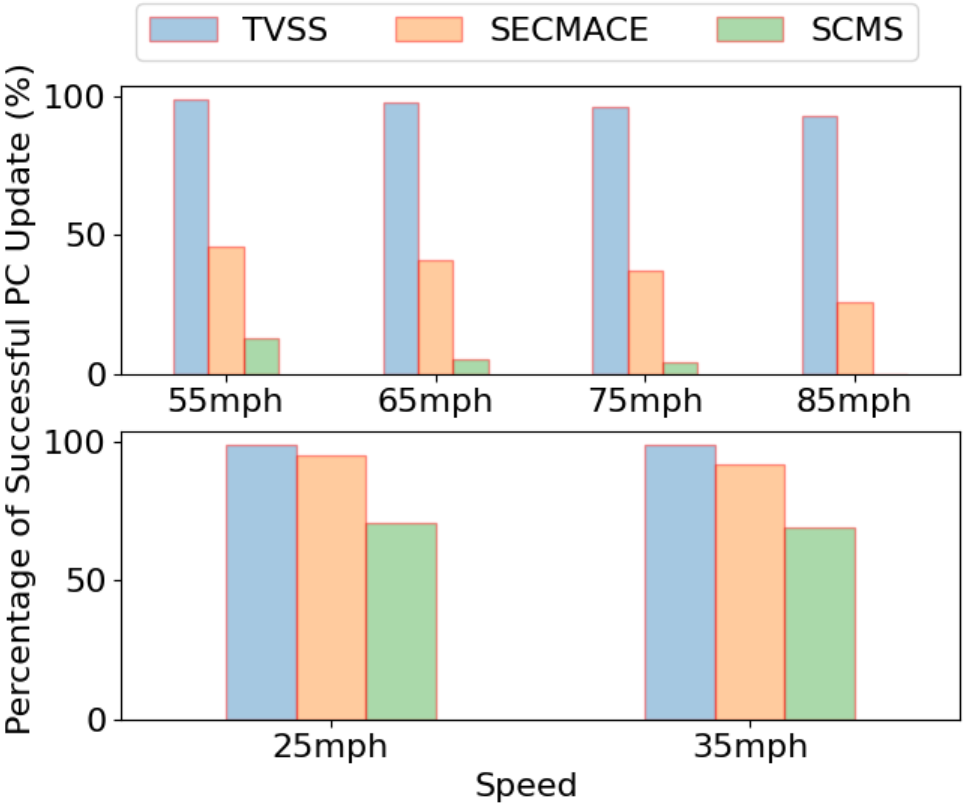}
\caption{Success ratio of vehicles refreshing a PC via all VPKIs under highway speeds (top) and in-city speeds (bottom).}
\label{fig:Percentage of Success Of Pc Gen}
\end{figure}

\vskip 3mm
\textbf{Evaluation 3: PC Renewal Success Ratio.} In this evaluation, we want to measure the ability of various VPKI systems to issue a PC under the constraint of a vehicle coverage time when it drives by an RSU. We test the three VPKIs under two scenarios: 1) highways with speeds of 55mph-85mph and 2) in-city with speeds of 25mph-35mph. The results are reported in Figure~\ref{fig:Percentage of Success Of Pc Gen} (top). From the figure, we can see that for the case of a regular highway speed of 55mph, PC refreshes can be completed successfully with a ratio of 99\% for \name{}, 46\% for SECMACE and 13\% for SCMS; this means that \name{} achieves 2.15x and 7.62x better reliability compared to SECMACE and the standard VPKI SCMS, respectively. For the the worst case of a CV driving on the highway at 85mph, PC refresh success ratio are 93\% for our proposed system \name{} and 26\% for SECMACE. For SCMS, however, it was unable to refresh PCs at such small RSU-OBU contact time. This shows that \name{} improves performance over SECMACE by 3.58x and is a reliable under such harsh conditions.

Furthermore we test all VPKI protocols on a city street under low speeds (i.e., in-city speeds), namely 25mph and 35mph, and show their PC issuance success ratio in Figure~\ref{fig:Percentage of Success Of Pc Gen} (bottom). The results show that when vehicles are driving at 25mph, the PC generation success ratios are 99\% for \name{}, 95\% for SECMACE and 71\% for SCMS. While \name{} outperforms SECMACE and SCMS, it is clear that cloud-based VPKI protocols are acceptable for slow moving vehicles but not for fast velocities at highways, which usually constitute most of the vehicle's trip. One can see that SECMACE and SCMS have a difference in their performance although both rely on the cloud; this is because SECMACE has a more efficient design and divides the work needed to issue a new PC between the cloud and an RSU while SCMS completely relies on the cloud. Our protocol, on the other hand, utilizes minimal cryptographic operations for the PC generation issuance process  and moves it completely to the edge (i.e., RSU), which allows for this improvement.

\cut{
\textbf{Evaluation 4: Delay Tolerance of SCMS and \name.} 
In this evaluation, we, as shown in Figure~\ref{fig:breaking points}, plot the latency required to generate a PC in SCMS and our proposed system under various server loads.
On the y-axis, we mark the median and average coverage times with the RSU for vehicles driving at high speeds (i.e., 75-90mph) to better understand the implications of our results. In order to compare the delay tolerance of SCMS and our proposed system, we pinpoint the maximum server delays a vehicle driving on the highway can tolerate to complete the PC generation procedure successfully in both systems. SCMS shows that it can tolerate  up to $\sim$1 second of server delay after which the vehicle loses connectivity with the RSU, which results in PC generation failure. Our system, on the other hand, can tolerate  increased server delay of 2.29x compared to SCMS, which makes our system a reliable alternative optimized for use by vehicles on freeways. This is because our system  utilizes the edge computing paradigm by completely bringing PC generation closer to the vehicle, which in turn reduces the network latency by 300 times (a significant reduction).
\begin{figure}[t]
\centering
\includegraphics[width=\linewidth,keepaspectratio]{Images/e2e latency scms vs ours.pdf}
\caption{Delay tolerance to update a PC in SCMS vs. \name.}
\label{fig:breaking points}
\end{figure}
}

\vskip 3mm
\textbf{Evaluation 4: PCRL Size.}
In this evaluation, we discuss the feasibility of pseudonym certificate revocation lists (PCRLs) used in \name\ and other VPKI systems. Recall that PCRLs hold pseudonym identities of revoked vehicles and are used to notify and protect CVs from malicious vehicles. In all calculations of other VPKI protocol PCRLs, we consider the hash chain optimizations used in SCMS to reduce the PCRL size while in transit. \name{} on the other hand, does not need to use these optimizations as each OBU can have a single PC at any time interval. In order to provide a reasonable projection of PCRL sizes in VPKI, we consider the revocation ratios in another similar domain, namely the internet; ~\cite{smith2020let} shows that the regular revocation ratio in the internet is $\sim$1.29$\%$ of the whole PKI system. Because there might also be cases of mass revocation such as the cyber attack in 2015 which allowed hackers to disable vehicles and triggered Chrysler to recall 1.4 million vehicles~\cite{Cyberhac5:online}, we also consider mass revocation with a ratio of $5\%$. Considering the US scenario with a total of 350 million vehicles, this would result in a PCRL of $\sim$177 MB in case of regular revocation and $\sim$684 MB in case of mass revocation as shown in Table~\ref{tab:PCRL}, which both require high bandwidth. In order to reduce the bandwidth requirements of downloading a PCRL, VPKIs additionally consider utilizing delta PCRLs, where the PCRL is incrementally updated so that the vehicle only downloads the newly revoked vehicles (i.e., delta PCRL) when it gets network connectivity with the RSU/backend (e.g., weekly/daily); nonetheless, the PCRL storage requirement at the vehicle stays the same as mentioned earlier. We also divide the size of the entire US PCRL size by the number of weeks and days to compute sizes of the weekly/daily delta PCRLs, respectively, as shown in Table~\ref{tab:PCRL}.

In order to reduce the PCRL size, VPKI systems also suggest dividing the revoked certificates on different PCRLs based on some common factor, such as region of revocation (e.g., a state)~\cite{Brecht2018SCMS}, so that a vehicle only downloads and keeps a relevant PCRL to it. In Table~\ref{tab:PCRL}, we show the anticipated size of PCRLs in the top four states in the US based on number of vehicles in each state. All states PCRLs require high bandwidth and storage as shown in Table~\ref{tab:PCRL}. Notice that this approach is not secure because it can still allow a revoked vehicle to maliciously participate in the system outside the region of revocation without being detected as long as the PCs are valid (e.g., SCMS PCs are valid for a week); this can degrade the safety and efficiency of the vehicular environment. 

\begin{table}[t]
\centering
  \begin{tabular}{lSS}
    \toprule
    \multirow{2}{*}{Region}  &
     \multicolumn{2}{c}{\makecell{PCRL Size (MB)}} \\ 
       & {\makecell{Regular Revocation\\(1.29\%)}} & {\makecell{Mass Revocation\\(5\%)}} \\
      \midrule
    \makecell[l]{Entire US} & {176.4} & {683.6}   \\
    California & {19.9} & {77.3}     \\
    Texas & {14.7} & {56.9}   \\
    Florida & {11.4} & {44.1}  \\
    New York & {7.3} & {28.2}    \\
    Delta PCRL (weekly) & {3.4} & {13.1}   \\
    Delta PCRL (daily) & {0.49} & {1.87}   \\
    Local PCRL & {0.04} & {0.16}   \\
    \bottomrule
  \end{tabular}
  \caption{PCRL types and sizes.}
  \label{tab:PCRL}
\end{table}

\name{}, on the other hand, ties each PC to a very small region (i.e., an area between RSUs), which allows for significantly reduced PCRLs, without having the aforementioned vulnerability. To predict a realistic size of a local PCRL, we consider the total number of vehicles that can fill up an area between RSUs in a highway. Particularly, we consider the real Wyoming deployment of RSUs in which RSUs are at most 20 miles apart~\cite{WYDOT:online}. Our results show that our technique of local PCRLs allows them to be extremely small even in case of mass revocation ($\sim$160KB), which are 13x smaller than delta PCRLs (i.e., the best SCMS choice) which eventually improves the security of the system. Please note that SCMS can not be adapted to tie PCs to small regions as this would violate anonymity and allow for tracking by the backend.

\begin{figure}[t]
\centering
\includegraphics[width=.8\linewidth,keepaspectratio]{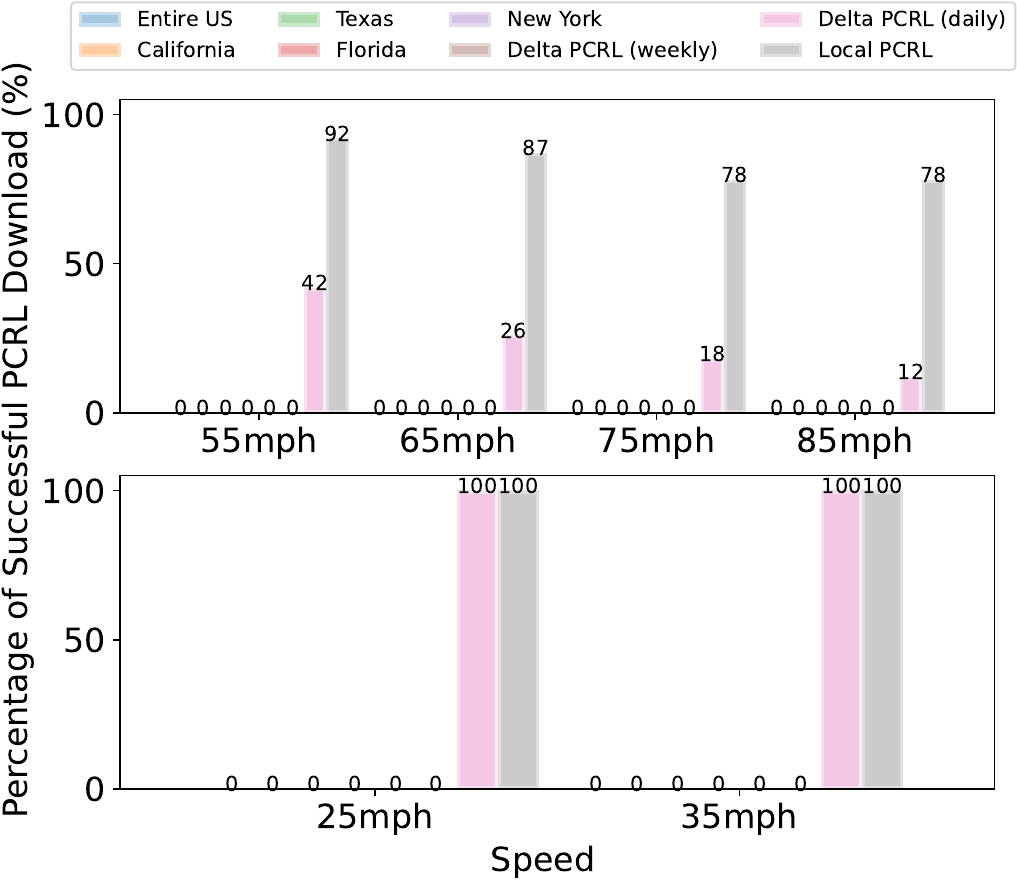}
\caption{The success ratio of downloading a complete PCRL under \emph{regular} revocation for various PCRL sizes in highway speeds (top) and in-city speeds (bottom).}
\label{fig:Percentage of Success Of pcrl download}
\end{figure}

\begin{figure}[t]
\centering
\includegraphics[width=.8\linewidth,keepaspectratio]{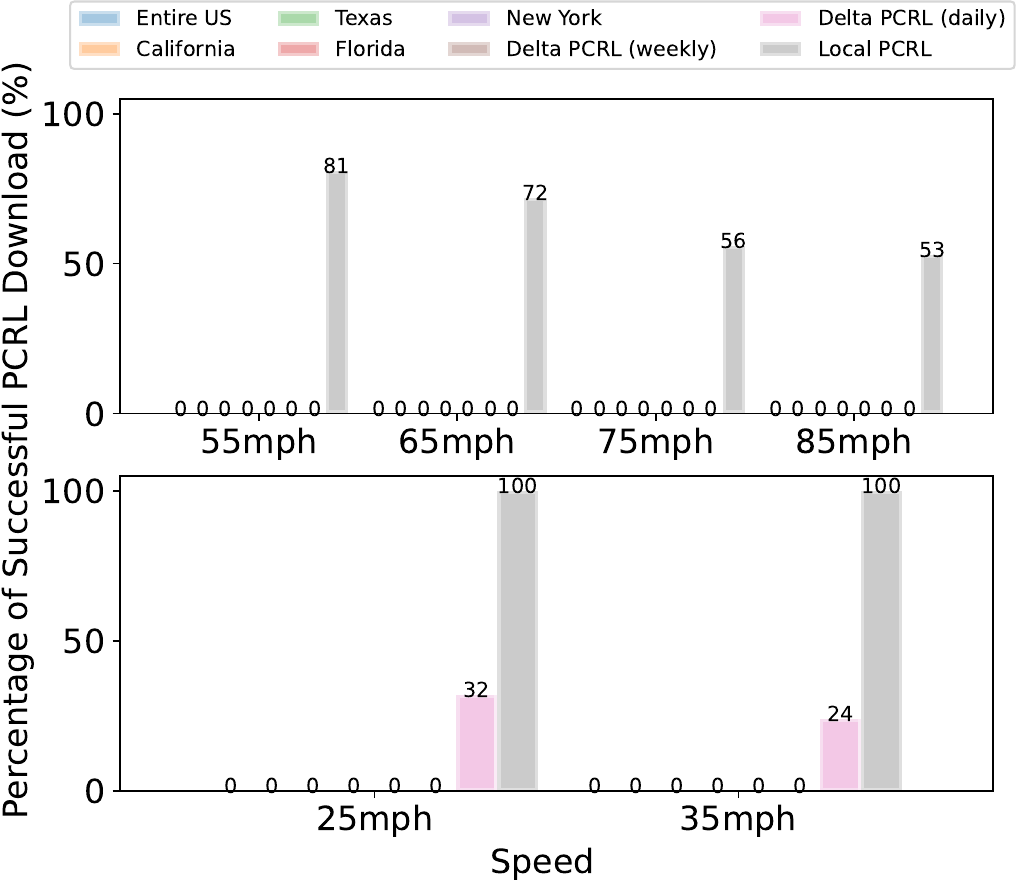}
\caption{The success ratio of downloading a complete PCRL under \emph{mass} revocation for various PCRL sizes in highway speeds (top) and in-city speeds (bottom).}
\label{fig:Percentage of Success Of pcrl download mass}
\end{figure}

\vskip 3mm
\textbf{Evaluation 5: Ratio of Successful Download of PCRL.} 
In this evaluation, we conduct a second set of field experiments to show the ratio of success to download PCRLs by CVs while driving on highways and in the city when passing by RSUs for both cases of regular revocation (Figure~\ref{fig:Percentage of Success Of pcrl download}) and mass revocation (Figure~\ref{fig:Percentage of Success Of pcrl download mass}).
Note that these results were collected using the baseline DSRC data rate of 6Mbps as specified in~\cite{adams2016development}. The PCRL sizes considered in this experiment are the PCRLs that cover the entire US, delta PCRLs (that are weekly/daily updated), specific states PCRLs, and small regions (i.e., our novel local PCRLs). The size of each PCRL type is shown in Table \ref{tab:PCRL}. From the experiment results, we observe that in both regular and mass revocations, OBUs are not able to download the whole nation-wide, states, and weekly updated PCRLs due to their massive size, which leaves OBUs unable to detect malicious OBUs; thus, these PCRL types are not feasible solutions as they make OBUs become oblivious of revoked and malicious OBUs. Note that for all PCRL types, we assume RSUs already have them stored locally and thus OBUs download them directly from the RSUs without the need to contact cloud servers so as to reduce network latency. Also note that alternatives to PCRLs such as certificate status inquiry protocols requiring communication with the CA or some central servers (e.g., OCSP) are not feasible as OBUs use DSRC only and do not always have contact with RSUs.

Under regular revocation (Figure~\ref{fig:Percentage of Success Of pcrl download}); for the regular highway speed of 55mph, only $42\%$ can download daily delta PCRLs successfully, leaving $58\%$ of legitimate OBUs vulnerable to bogus messages sent by malicious OBUs. On the other hand, our local PCRLs show promising results as $92\%$ of the OBUs are able to download them. For the highest highway speed of 85mph, only $12\%$ can download daily delta PCRLs, leaving $88\%$ of legitimate vehicles unaware of malicious vehicles. Our local PCRLs show promising results as $78\%$ of OBUs can download them at this high speed. This improves the safety and efficiency of the vehicular environment as CVs can quickly detect malicious messages broadcast by revoked OBUs. For in-city speeds (i.e., 25mph and 35mph), the results show that $100\%$ success ratio of downloading local PCRLs and daily delta PCRLs under regular revocation, as expected. 
 
Due to the large scale of VPKI, we additionally consider the effect of mass revocation on the system (Figure~\ref{fig:Percentage of Success Of pcrl download mass}). Our experimental results show that in case of mass revocation (i.e., $5\%$ revocation rates), OBUs now cannot download daily delta PCRLs at highway speeds (55mph-85mph), which indicates unscalability of this technique under mass revocation. In contrast, \name{} shows substantial tolerance to such mass revocation cases since $81\%$ and $53\%$ of vehicles are able to download local PCRLs at 55mph and 85mph, respectively. For in-city speeds under mass revocation, local PCRLs achieves $100\%$ download success ratio while only $32\%$ of OBUs can download the alternative daily delta PCRLs.  This makes local PCRLs a scalable solution, that improves the security of CVs.

\vskip 3mm
\textbf{Evaluation 6: RSU Scalability.}
Since \name\ proposes to move the whole workload to generate PCs to the edge network,  we, in this experiment, examine the scalability of current road-side equipment (i.e., RSUs) to handle such workloads. This is important because we want to make sure that we do not have to incur extra costs by upgrading current RSUs or otherwise end up with a fragile deployment. Specifically, we think about extreme situations where an RSU has to service many OBUs to refresh their PCs in a short time period. We also include SECMACE in our discussion of this experiment to show its RSU scalability (even though SECMACE does not rely on the RSU completely). As shown earlier in Figure~\ref{fig:Pc Gen Latency}, the latency to generate a PC using \name{} and SECMACE on a real RSU are 1.08ms and 6.95ms, respectively. This latency is mostly dominated by cryptographic operations. Please note that, while online/offline cryptographic digital signature schemes (e.g., \cite{Mss2021Abdul, sharma2017pf, johnson2001elliptic}) can be utilized here to offer faster signature generation in the critical online time where vehicles are temporarily in range of an RSU, our results and discussion here do not consider them. 

\begin{figure}[t]
\centering 
\includegraphics[width=.8\linewidth,keepaspectratio]{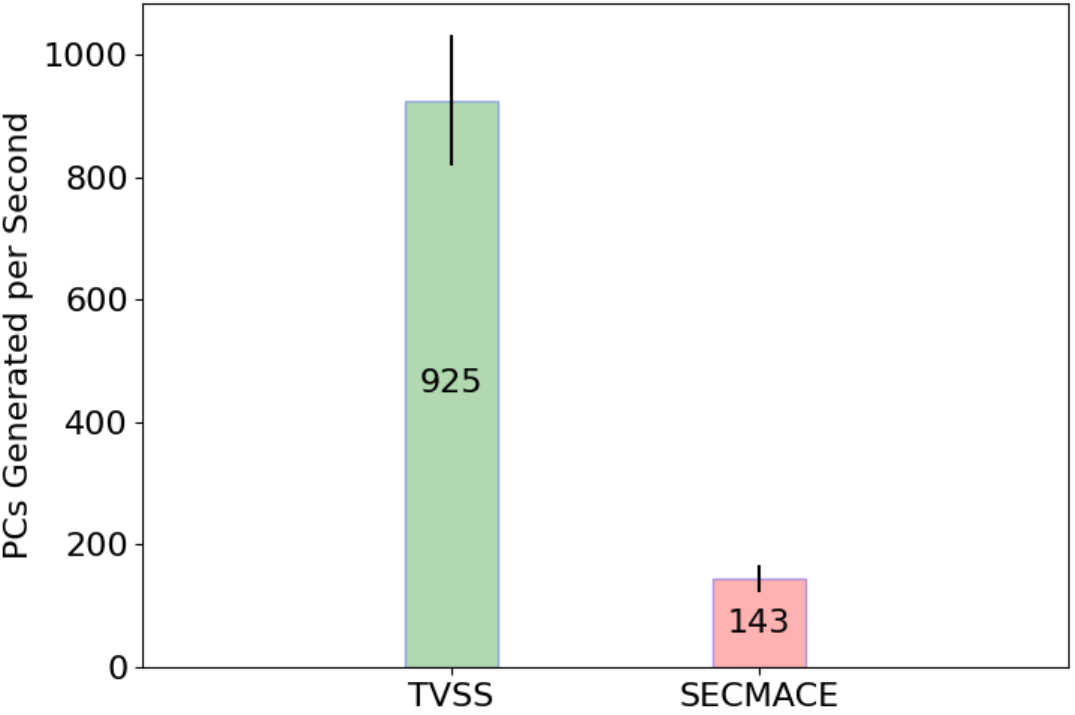}
\caption{Number of PCs generated per second by an RSU.}
\label{fig:rsu scalability}
\end{figure}

Consequently, Figure~\ref{fig:rsu scalability} shows that one RSU could generate 925 PCs per second for \name{} and 143 PCs per second for SECMACE. To put this into context, we consider a busy interstate and estimate the number of vehicles in a single RSU coverage area. Considering the radius of 150 meters as specified by US DoT~\cite{perry2017dedicated}, an interstate with 8 lanes could contain at most 500 vehicles.  Our results show that the RSU is able to service 1.85x the vehicles at a busy interstate when using \name{}. This indicates that RSUs have the enough computational capability to generate PCs without having to upgrade them and incur extra costs. It also shows that \name{} presents a scalable solution for VPKI. SECMACE, however, does not scale as it can service only $28.6\%$ of PC generation requests. 

\begin{figure}[t]
\centering 
\includegraphics[width=\linewidth,keepaspectratio]{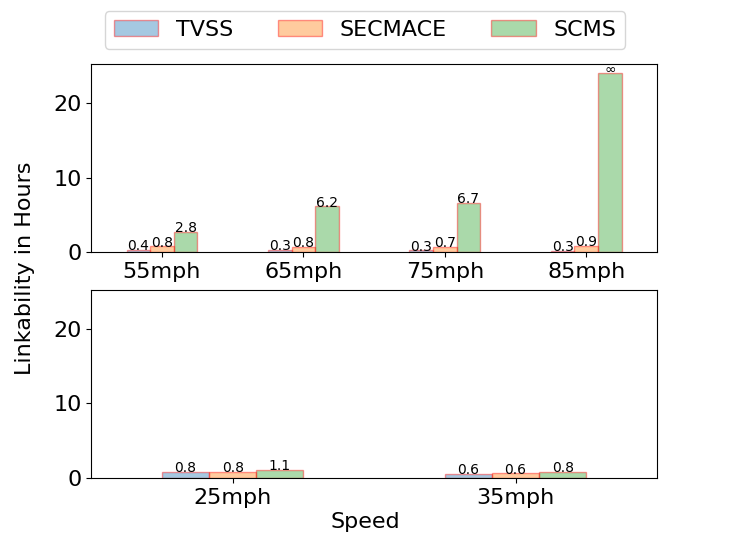}
\caption{Vulnerability window to track vehicles in \name{}, SECMACE and SCMS.}
\label{fig:linkability window}
\end{figure}

\vskip 3mm
\textbf{Evaluation 7: Linkability Window in VPKIs.}
  Recall that the unreliability of the CV wireless networks causes a CV to reuse the same PC for longer periods if the CV is unable to renew the PC when driving by an RSU. We, in Figure~\ref{fig:linkability window}, show the linkability windows that curious vehicles/authorities have to link messages of a single vehicle when using different VPKIs for two different scenarios, namely in-city and highways; this eventually leads to other vehicles/authorities tracking vehicle driving activities.  In order to provide realistic linkability windows, we consider the aforementioned Wyoming deployment in which RSUs are placed 20 miles apart. Intuitively, we expect that the faster the CV drives, the faster it reaches the next RSU (i.e., smaller duration between two RSUs) and updates its PC; this allows PC validity $T$ to become smaller and thus the vulnerability window. As expected in our system, the results show that the faster the vehicle drives, the smaller the linkability window becomes; this is because CVs successfully refresh PCs whenever in vicinity of an RSU in our system. In contrast, SCMS and SECMACE show a similar pattern but a sharp increase in the vulnerability window when CVs drive on the highway due to the increased failure ratio of refreshing PCs at high speeds. In particular, SCMS and SECMACE show that a single CV driving on the highway at 65mph can be tracked for 6.2 hours and 0.8 hours, respectively, whereas \name{} significantly reduces that vulnerability window to 18 minutes (i.e., $\sim$22.5x and 2.7x reduction in linkability window compared to SCMS and SECMACE, respectively). This makes it harder for adversaries to track CV driving activities due to the limited linkability window. Therefore, \name{} can considerably decrease $T$, which improves the privacy of the overall system.

\section{Related Work}
\label{sec:related}


In this Section, we discuss closely related work.

\noindent\textbf{Early VPKI Designs:} In its early designs \cite{wiedersheim2009sevecom,papadimitratos2008secure,bissmeyer2014preparing}, the VPKI primarily consisted of a CA that issues enrollment certificates ($\ltc$s) to vehicles and later issues PCs for V2V communication. While this protects a CV from tracking from other CVs, the CA can still track all CVs.

\noindent\textbf{Separation of Duties in a VPKI:} At later iterations, the design of the VPKI separated the PC issuance tasks from CA and assigned them to a different VPKI entity named Pseudonym Certification Authority (PCA). The main motivation of this separation is to prevent a single entity from linking an $\ltc$ to its corresponding PC.
The US department of transportation (US DoT) has mandated a VPKI system called Security Credential Management System (SCMS)~\cite{Brecht2018SCMS, USDOTSCMS}. Similarly, the European Union (EU) has mandated a VPKI system called cooperative intelligent transport systems (C-ITS) certificate management system (EU CMSS)~\cite{certpolicy2018cits,etsi2021102940, etsi2021102941}. In these systems, vehicles get fresh PCs from the PCA only after PC requests are validated by the CA using vehicles' $\ltc$s. The separation of duties helps prevent the PCA from revealing vehicles' $\ltc$s and the CA from revealing PCs information (e.g., public/verification keys).
Furthermore, the authors in \cite{alexiou2013vespa, gisdakis2013serosa,bissmeyer2013copra,khodaei2016evaluating} have proposed a \textit{ticketing scheme} where a vehicle requests a \textit{PC-ticket} from the CA then uses this ticket to acquire a PC from a PCA. The authors in \cite{schaub2010v} further improved the concept of tickets and introduced what they call a V-Token, an encrypted PC-$\ltc$ linkage information which is embedded inside every PC and can only be decrypted by the collaboration of several VPKI entities using a threshold encryption scheme. V-Token also utilizes a blind signature scheme to prevent the CA from peering into the content of tokens but at the expense of revealing (hence discarding) some of the tokens in order to provide a probabilistic authentication for the remaining tokens.

\noindent\textbf{Fault-Tolerance in a VPKI:} As a different improvement, the authors in \cite{khodaei2018secmace} have proposed SECMACE, a VPKI architecture that \textit{splits} the CA into multiple CAs where each CA is responsible of a manageable set of vehicles (usually bounded by the same geographical region). The authors in \cite{tesei2018iota} have proposed an additional improvement over SECMACE by introducing IOTA-VPKI, a Distributed Ledger Technology (DLT) implementation based on Direct Acyclic Graph (DAG). The main purpose of this improvement is to prevent a \textit{single-point-of-failure} whenever one of the CAs goes down. The authors of SECMACE have further improved their VPKI design by implementing it as a \textit{cloud-based} solution named VPKI as a Service (VPKIaaS) \cite{khodaei2019scaling} making use of many cloud-based paradigms such as server migration and resource expansion.

\underline{Drawbacks:} All of the previously mentioned VPKI designs heavily rely on backend services that are located behind a network infrastructure. Furthermore, every PC request needs at least two separate roundtrips across the network (vehicle-CA and vehicle-PCA roundtrips).

\noindent\textbf{Hierarchical VPKI Certification:} As a notable work, the authors in \cite{ma2008pseudonym} have proposed a VPKI design that is similar to ours. Instead of obtaining PCs, a vehicle obtains Long-Term PCs (LPC) from a PCA that can be valid up to months. The vehicle then obtains the regular short-termed PCs using an LPC from another new entity in the VPKI named Road Authority (RA) which is responsible of issuing PCs for a specific geographical region. A single RA typically controls multiple RSUs that cover the region of the controlling RA.

\underline{Drawbacks:} With such a hierarchical design, there is a high probability that a vehicle would request PCs from the same RA using the same LPC because of 1) the long-term validity time of LPC entails a bigger exposure of its signature, and 2) the repetitive routine of the daily commutes of drivers which make them interact with the same RA on a daily basis (e.g. from home to work and vice versa). This situation would enable an RA to link multiple PCs to the same vehicle.
\section{Conclusion}
We realize that CVs only have short unreliable connection times with RSUs, the main gateway for internet. This causes cloud-based VPKIs to likely fail to refresh vehicle PCs due to high network latency. Thus, we propose a novel token-based VPKI, called \name{}, that addresses the shortcomings of other VPKIs. \name{} makes use of the edge-computing paradigm, where the PC issuance tasks are pushed to the RSUs. This allows CVs to obtain PCs with low network latency without violating anonymity or unlinkability properties.  Our field experiments on the street shows that \name{} is reliable even under high driving speeds as opposed to its alternatives;  \name{} can successfully generate PCs for vehicles with a $99\%$ chance whereas that ratio drops to $42\%$ for the best VPKI alternative. \name{} also provides pseudonym certificate revocation lists (PCRLs), that are 13x smaller compared to traditional PCRLs.



%

\bibliographystyle{plain}
\bibliography{TVSS}

\appendix
\subsection{Cryptographic Preliminaries}
\label{sec:crypto prelim}
In this section, we describe the basic cryptographic systems that are utilized in VPKIs (suhc as SCMS and \name{}), namely encryption, digital signatures, decisional Diffie-Hellman (DDH), and hash chains.

\vskip 3mm\noindent\textbf{Encryption Schemes. }
Encryption schemes are used in cryptography to ensure confidentiality of the information being sent over a network. It can be formally defined as a set of three algorithms $(\keygen, \encrypt, \decrypt)$. The algorithms $\keygen(1^n)$ outputs the key pair $(\enckey,\deckey)$, $\encrypt(\msg,\enckey)$ outputs a cipher text $\ct$ and $\decrypt(\deckey,\ct)$ outputs the message, $\msg$, by using the decryption key $\deckey$. It satisfies the properties of correctness and the security. Correctness says that for all messages, if $(\enckey,\deckey)\leftarrow\keygen(1^n)$, then $\decrypt(\deckey,\encrypt(\msg,\enckey))=\msg$. Informally, the security property is that no efficient adversary can decrypt the ciphertext without having access to the decryption key.

\vskip 3mm\noindent\textbf{Digital Signatures.}
Digital signatures are basic schemes used in cryptography and used for authentication purposes. A signature scheme can be formally defined as a set of three algorithms $(\keygen, \sign, \verify)$. The algorithms $\keygen(1^n)$ outputs the key pair $(\vk,\sk)$, $\sign(\msg,\sk)$ outputs a signature $\sigma$ and $\verify(\vk,\msg,\sigma)$ checks whether the signature is valid or not.

Additionally, the correctness and security of the digital signatures must also hold. Correctness says that for all messages $\msg$, if $(\vk,\sk)\leftarrow\keygen(1^n)$ and $\sigma\leftarrow\sign(\msg,\sk)$, then $\verify(\vk,\msg,\sigma)=1$.  Intuitively, security demands that without possession of the secret key, no adversary can produce a valid signature for a new message.

\vskip 3mm\noindent\textbf{Decisional Diffie-Hellman.}
We define the decisional Diffie-Hellman (DDH) problem in a cyclic group $G$ with a generator $g$. It is defined as follows. Given a tuple $(g,g^x,h,h')$ where $x$ is a random exponent, $h\in G$ is random and $h'$ either equals $h^x$ or else is a random group element, decide which is the case. In DDH based encryption schemes, the private/public key pair $(x, g^x)$ corresponds to the decryption and encryption keys ($\deckey,\enckey$), respectively. Similarly, the key pair $(x, g^x)$ corresponds to the signature and verification keys ($\sk,\vk$), respectively, in digital signature schemes.

\begin{figure}[t]
\centering
\includegraphics[width=\linewidth,keepaspectratio]{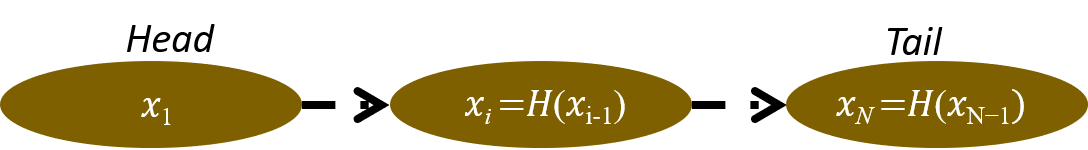}
\caption{A hash chain.}
\label{fig:Hash Chain}
\end{figure}

\vskip 3mm\noindent\textbf{Hash Chains.}
Hash chains are important primitives which appear in authentication schemes. Formally, we can define a hash chain as follows. Given a hash function $H$, a \emph{hash chain} is a list of vertices $\{v_1,\dots,v_N\}$ where each $v_i$ is labeled by a string $x_i$ such that $x_{i+1}=H(x_i)$ holds for all $i$ (Figure~\ref{fig:Hash Chain}).  Since it is computationally infeasible to invert $H$, hash chains acquire a notion of direction.  The first vertex $v_1$ is called the \emph{head} of the chain, $v_N$ is the \emph{tail}. The property that is required is as follows. Given $x_i$, one can compute $H(x_i)=x_{i+1}$ efficiently, but not the other way because of the hardness assumption.

\subsection{Security Proof}
\label{sec:secproof}
Assume for contradiction that an efficient adversary $\mathcal{A}$ wins the VPKI game with non-negligible advantage.  We construct another efficient adversary $\mathcal{B}$ which breaks the security of the signature scheme $({\sf Gen},{\sf Sign},{\sf Verify})$ with related advantage.  The first thing $\mathcal{B}$ does is it guesses how $\mathcal{A}$ will try to win the VPKI game; it will proceed differently in the three cases.

Suppose first, that $\mathcal{B}$ decides $\mathcal{A}$ will win the forgery version of the VPKI game.  In this case, $\mathcal{B}$ begins playing as the adversary in the security game for $({\sf Gen},{\sf Sign},{\sf Verify})$.  Specifically, the signature game challenger $\mathcal{C}$ draws $({\sf vk},{\sf sk})\sim{\sf G}$ and sends ${\sf vk}$ to $\mathcal{B}$.  Then $\mathcal{B}$ invokes $\mathcal{A}$ and plays as the challenger in the VPKI game.  $\mathcal{B}$ picks a random vehicle $v$ and a random PC for $v$ and uses ${\sf vk}$ as the public verification key portion of the PC; all other queries of $\mathcal{A}$ are answered honestly by $\mathcal{B}$, just as the honest challenger would answer them.  If $\mathcal{A}$ does indeed decide to win the forgery version of the game, and moreover if $\mathcal{A}$ intends to use the selected vehicle $v$ and the selected PC to produce the forgery then $\mathcal{B}$ proceeds, otherwise $\mathcal{B}$ aborts.  In case $\mathcal{B}$ proceeds, $\mathcal{B}$ simply forwards $\mathcal{A}$'s forged message to $\mathcal{C}$.  It is clear that $\mathcal{B}$ wins the forgery game whenever $\mathcal{A}$ wins the forgery version of the VPKI game using the vehicle $v$ and the selected PC.  Thus, if $\mathcal{A}$ has a non-negligible advantage of winning the forgery version of the VPKI game, then $\mathcal{B}$ has non-negligible probability of breaking the signature scheme.

Now, let us suppose that $\mathcal{A}$ has a non-negligible advantage of breaking either the unlinkability or the anonymity versions of the VPKI game.  We derive an information theoretic contradiction in this case by showing that $\mathcal{A}$ has the ability to predict a random bit with positive advantage.  The key point here is that the PCs are completely independent of the vehicle and of each other and so there is no way $\mathcal{A}$ can win the anonymity or unlinkability branch of the game with probability better than $1/2$.  We prove this for anonymity, the case of unlinkability is similar.  Consider an adversary $\mathcal{B}$ who emulates the VPKI challenger in all ways except that it chooses two random vehicles $v_0$ and $v_1$ and it generates two PCs: ${\sf P}_0$ and ${\sf P}_1$.  Neither PC is associated yet with either vehicle.  Now $\mathcal{B}$ flips a coin; if heads, $\mathcal{B}$ associates ${\sf P}_b$ with $v_b$ for $b=0,1$; if tails $\mathcal{B}$ associates ${\sf P}_b$ with $v_{1-b}$ for $b=0,1$.  Now, if $\mathcal{A}$ decides to win the anonymity branch of the VPKI game using vehicles $v_0$ and $v_1$, then $\mathcal{B}$ sends ${\sf P}_0$ to $\mathcal{A}$ (if $\mathcal{A}$ decides to win a different branch, or decides to use different vehicles, then $\mathcal{B}$ outputs a random bit.  Now, note that if $\mathcal{B}$'s coin landed on $b$, then $\mathcal{A}$ needs to return $b$ to win.  However, as $\mathcal{B}$'s challenge is independent of $b$, the probability that $\mathcal{A}$ returns $b$ is exactly $1/2$.  Therefore, it is not possible for $\mathcal{A}$ to have an advantage in the anonymity branch of the VPKI game.  The same argument shows that $\mathcal{A}$ cannot have any advantage in the unlinkability branch.

So in summary, it must be that if $\mathcal{A}$ has a non-negligible advantage in winning the VPKI game, then this advantage must lie in the forgery branch of the game.  However, as shown above, in this case we can design an efficient adversary $\mathcal{B}$ which breaks the security of the underlying signature scheme.

\end{document}